\documentclass[runningheads]{svmult}

\usepackage{makeidx}   
\usepackage{graphicx}  
\usepackage{subeqnar}  
\usepackage{multicol}  
\usepackage{physprbb}  
\makeindex             

\begin{document}
\title*{Heavy Quarks on the Lattice}
\toctitle{Focusing of a Parallel Beam to Form a Point
\protect\newline in the Particle Deflection Plane}
%
%
\titlerunning{Focusing of a Parallel Beam}
%
\author{Craig McNeile\inst{1}
%
%
\institute{
Department of Mathematical Sciences, 
University of\ Liverpool, L69 3BX, UK}
}

\maketitle              

\begin{abstract}
  I review the basic ideas behind lattice QCD calculations that
  involve charm and bottom quarks. I report on the progress
  in getting the correct hyperfine splitting in charmonium from
  lattice QCD.
  Some of the basic technology
  behind numerical lattice QCD calculations is explained by studying
  some specific examples: computation of the charm quark mass, and the
  calculation of $f_B$.
\end{abstract}

\section{Introduction}

The B factories at KEK and SLAC are producing a wealth of new data on
the decays of the B meson.  One of the main goals of the current heavy
flavor program is to check the CKM matrix formalism by measuring
the matrix elements with sufficient accuracy.  To convert the
experimental data into information about the quarks requires the
accurate computation of hadronic matrix elements.  The best (and some
would say only) way of computing the required matrix elements is to
use lattice QCD.


In this paper I will explain the generic features of lattice QCD
calculations that involve heavy quarks.  
Heavy quark lattice calculations share many common features to continuum
calculations, such as matching to effective field theories. 
However, the more general formalism of lattice
QCD allows a richer set of tools beyond 
just using perturbation theory.  As most
lattice QCD calculations share generic features, I will work through an
example of computing the charm quark mass to show the important parts
of the calculation. I will then discuss the calculation of the $f_B$
decay constant.
I assume the reader is already familiar with the basic ideas of heavy
quark effective field theory~\cite{Neubert:1994mb,Manohar:2000dt}.

The latest results on heavy quark physics from the lattice
are reported in the reviews at the annual lattice
conference~\cite{Draper:1998ms,Bernard:2000ki,Ryan:2001ej}.  The
contents of the proceedings of the lattice conference have been put on
hep-lat for the past couple of
years~\cite{Davies:1998ur,Mueller-Preussker:2002cp}
The longer reviews by Kronfeld~\cite{Kronfeld:2002pi}, 
Davies~\cite{Davies:1997hv,Davies:2002cx} and
Flynn and Sachrajda~\cite{Flynn:1998ca} contain
other perspectives on heavy quarks on the lattice.
Gupta gives a general overview of lattice QCD~\cite{Gupta:1997nd}.
There have been recent (political) developments to set up a working
group on producing a ``particle data table'' for lattice
QCD results~\cite{Flynn:2002yu}.

\section{A brief introduction to numerical lattice QCD} \label{se:intro}

Most lattice QCD calculations start from the calculation
of the correlator $c_{ij}(t)$ defined in terms of the 
path integral as:
\begin{equation}
c_{ij}(t)
=
  \frac{1}{Z}\int dU\int d\psi \overline{d\psi
  }O(t)_{i}O(0)_{j}^{\dagger }e^{-S_{F}-S_{G}}
\label{eq:PATH}
\end{equation}
where $S_F$ is the action of the fermions and $S_G$
is the action of pure gauge theory.

\begin{figure}[b]
\begin{center}
\includegraphics[width=.8\textwidth]{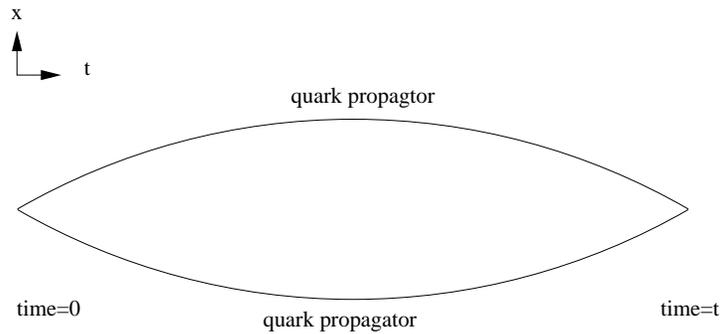}
\end{center}
\caption[]{Two point correlator}
\label{fg:twopoint}
\end{figure}

The path integral is regulated by the introduction of a four 
dimensional space-time lattice. 
A typical lattice volume would be $24^3 \;48$.
The path integral is evaluated
in Euclidean space for convergence.
The fermion action is 
\begin{equation}
S_F = \overline{\psi} M \psi
\label{eq:Faction}
\end{equation}
where $M$ is called the fermion operator, a lattice approximation to
the Dirac operator. The quadratic 
structure of the fermion action in equation~\ref{eq:Faction}
allows the integration over the fermion fields to be done explicitly.
\begin{equation}
c_{ij}(t)=\frac{1}{Z}\int dUO(t)_{i}O(0)_{j}^{\dagger
  }det(M)e^{-S_{G}}
\label{eq:PathOutGlue}
\end{equation}
The $det(M)$ term controls the dynamics of the sea quarks. The $O(t)$
operator controls the valence content of the state. For example an
operator ($O(t)$) for a B meson would be:
\begin{equation}
O(t) = \overline{b}(t)\gamma _{5}q(t)
\end{equation}
where $b$ and $q$ are operators for the bottom and light
quarks respectively.
The operators in the path integral are Wick contracted
to form a combination of quark propagators inside the 
path integral over the gauge fields.
\begin{equation}
c(t)=\frac{1}{Z}\int dU\overline{b}(t)\gamma
_{5}q(t)\overline{q}(0)\gamma _{5}b(0)
det(M)
e^{-S_{G}}
\label{eq:PathWick}
\end{equation}
The physical picture for the expression in~\ref{eq:PathWick}
is a B meson created at time 0 and propagating to time t where
it is destroyed. Figure ~\ref{fg:twopoint} shows the 
propagation of the light and heavy quark in the vacuum.

The path integral expression for the correlator in
equation~\ref{eq:PATH} is calculated using Monte Carlo techniques on
the computer. The ideas are sophisticated variants of the Monte Carlo
method used to compute integrals.

The algorithms, usually based on importance sampling, 
produce $N$ samples of the gauge fields on the lattice.
Each gauge field is a snapshot of the vacuum.
The QCD vacuum is a complicated structure. There
is a community of people who are trying to describe
the QCD vacuum in terms of objects such as a 
liquid of instantons (for example~\cite{Negele:1998ev}).
The correlator $c(t)$ is a function of the 
bottom ($M(U(i))_{b}^{-1}$ ) and light quark $M(U(i))_{q}^{-1}$ 
propagators averaged
over the samples of the gauge fields.
\begin{equation}
c(t) \sim  \frac{1}{N}\sum
_{i}^{N}f(M(U(i))_{b}^{-1},M(U(i))_{q}^{-1}) 
\label{eq:algorEXPLAIN}
\end{equation}
The
quark propagator is the inverse of the fermion operator.  In
perturbative calculations the quark propagator can be computed
analytically from the fermion operator.  In lattice QCD calculations the
gauge fields have complicated space-time dependence so the quark
propagator is computed numerically using variants of conjugate
gradient algorithms.

\begin{figure}[b]
\begin{center}
\includegraphics[width=.8\textwidth]{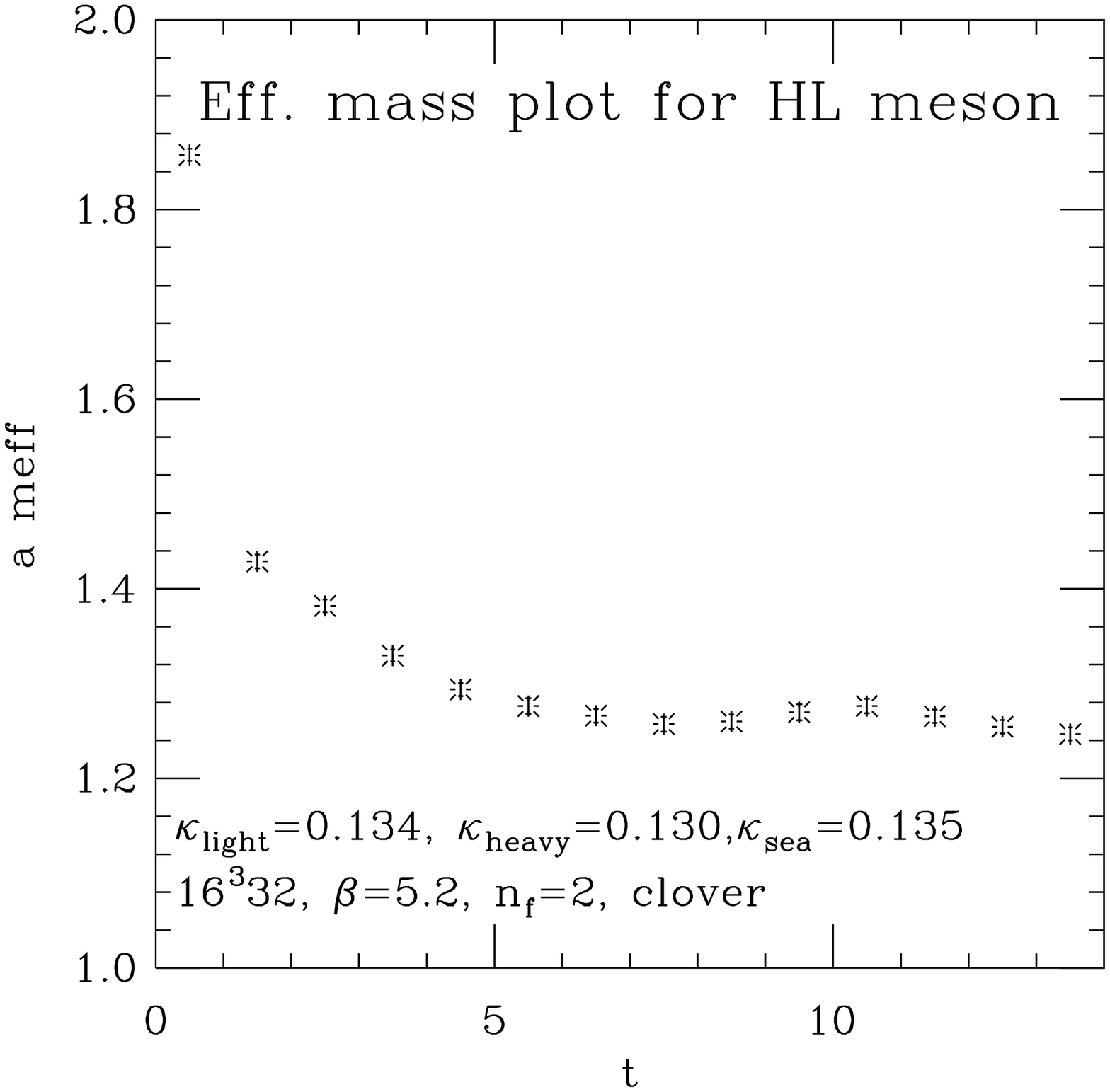}
\end{center}
\caption[]{Effective mass plot}
\label{fg:effective}
\end{figure}

The physics is extracted from the correlators by fitting
the correlator to a functional form such 
as~\ref{eq:FITmodel}.
\begin{equation}
c_{ij}(t)= c_{ij}^{0}e^{-m_{0}t} + 
           c_{ij}^{1}e^{-m_{1}t} +
           \ldots 
\label{eq:FITmodel}
\end{equation}
To visually judge the quality of the data,
the correlators are often displayed as effective
mass plots
\begin{equation}
m_{eff}(t) = log(\frac{c(t)}{c(t+1)})
\end{equation}
An example of an effective mass plot 
(using the data generated for~\cite{Maynard:2001zd})
is in 
figure~\ref{fg:effective}.

The computationally expensive part of lattice QCD calculations is
generating the samples of gauge fields. The most expensive
part of a lattice calculation is incorporating the 
determinant in equation~\ref{eq:PathOutGlue}.
The SESAM collaboration~\cite{Lippert:2002jm} estimated that the number
of floating point operations ($N_{flop}$) needed for $n_{f}$ =2
full QCD calculations as:
\begin{equation}
N_{flop} \propto (\frac{L}{a})^{5}(\frac{1}{am_{pi}})^{2.8}
\label{eq:cost}
\end{equation}
A flop is a floating point operation such as a multiplication
or addition. The value of $N_{flop}$ represents amount of
calculation required on the computer and even more importantly
the cost of the computer required.

In some sense equation~\ref{eq:cost} (or some variant of it) is the 
most important equation in numerical lattice QCD. To half the size
of the pion mass used in the calculations requires essentially 
a computer that is seven times faster.
Equation~\ref{eq:cost} is not a hard physical limit.
Improved algorithms or techniques may be cheaper. In fact the
``Asqtad'' fermion action 
designed by the MILC collaboration is already 
computationally cheaper~\cite{Gottlieb:2001cf}
than the cost estimates in equation~\ref{eq:cost}. 

The cost formula in equation~\ref{eq:cost}
is for the generation of the gauge 
configurations. 
Once the gauge configurations have been generated, correlators for
many different processes can be computed using some 
generalization of equation~\ref{eq:algorEXPLAIN}. This class
of calculation can be carried out on a farm of workstations.
The lattice QCD community are starting to create
publicly available source code~\cite{DiPierro:2001yu} 
and gauge 
configurations~\cite{McNeile:2000qm,Davies:2002mu}.

Table~\ref{tb:dynamicalPARAMS} shows the parameters of some recent
large scale unquenched calculations.
\begin{table}[tb]
\begin{center}\begin{tabular}{|c|c|c|c|c|}
\hline 
Collaboration&
$n_{f}$&
a fm&
L fm&
$\frac{M_{PS}}{M_{V}}$\\
\hline
\hline 
MILC~\cite{Bernard:1999xx}&
2+1&
0.09&
2.5&
0.4\\
\hline 
CP-PACS~\cite{AliKhan:2000mw}&
2&
0.11&
2.5&
0.6\\
\hline 
UKQCD~\cite{Allton:2001sk}&
2&
0.1&
1.6&
0.58\\
\hline 
SESAM~\cite{Glassner:1996xi}&
2&
0.08&
2.0&
0.56\\
\hline
\end{tabular}\end{center}
\caption{Typical parameters in recent unquenched 
lattice QCD calculations.}
\label{tb:dynamicalPARAMS}
\end{table}
It is not considered necessary to do lattice calculations with
physical light masses $M_{PS}/M_{V} = M_{\pi}/M_{\rho} \sim 0.18$.
The aim is calculate with light enough quarks so that chiral
perturbation theory can be used to make contact with the experiment.
Sharpe~\cite{Sharpe:1998hh} estimates that going as light as
$M_{PS}/M_{V}\sim 0.3$ may be necessary.

The high computational cost of the fermion determinant led
to development of quenched QCD, where the dynamics of the determinant 
is not included in equation~\ref{eq:PathWick}, hence the dynamics
of the sea quarks is omitted.  Until recently the majority
of lattice QCD calculations were done in quenched QCD.
When the dynamics of the sea quarks are 
included I will call the calculation unquenched.

Lattice QCD calculations produce results in
units of the lattice spacing. One experimental number must be 
used to calculate the lattice spacing from: 
\begin{equation}
a =
am_{latt}^{X} / m_{expt}^{X}
\end{equation}
As the lattice spacing goes to
zero any choice of $m_{expt}^{X}$ should produce the same lattice
spacing -- this is known scaling. Unfortunately, no calculations are 
in this regime yet. The recent unquenched calculations by the 
MILC collaboration~\cite{Gray:1990yh} may be close.

Popular choices to set the scale are the mass of the rho,
mass splitting between the S and P wave mesons in charmonium,
and a quantity defined from the lattice potential called $r_0$.
The quantity $r_0$ is defined by
$r_{0}$~\cite{Sommer:1994ce}.
\begin{equation}
r_{0}^{2} \frac{dV}{dr}\mid _{r_{0}} = 1.65 
\end{equation}
Many potential~\cite{Sommer:1994ce}
models predict $r_{0}$ $\sim $ 0.5 fm.
There is no perfect way to compute the lattice spacing. Although it may
seem a little strange to use $r_0$ to calculate the lattice
spacing, when
it is not directly known from experiment, there are problems with all
methods to set the lattice spacing.  For example, to set the scale
from the mass of the rho meson requires a long extrapolation 
in light quark
mass. Also it is not clear how to deal with the decay width of the rho
in Euclidean space.

\subsection{Fermion actions for light quarks}

The lattice QCD formalism has the quark fields on the nodes of the
lattice. The gauge fields are $SU(3)$ matrices and lie between the
nodes of the lattice. There are a variety of different ways of writing
a lattice approximation to the Dirac operator on the lattice.
Gupta~\cite{Gupta:1997nd} reviews the problems and possibilities of
fermion actions on the lattice.  Discussions between lattice gauge
theorists over which lattice fermion action is best must seem to
outsiders to have the flavour of fanatical religious discussions, with
the lattice ``community'' breaking into various sects, and accusations
of ``idolatry'' being flung around.  In the continuum limit all the
fermion actions should produce the same results. This is clearly a
good check on the results.

As a starting point I will consider the Wilson fermion
action.
\begin{equation}
S_{f}^{W}= \sum_{x}
(
\kappa \sum_{\mu }
\{
 \overline{\psi}_{x}(\gamma_{\mu }-1)U_{\mu }(x)\psi_{x+\mu }
-\overline{\psi}_{x+\mu}(\gamma_{\mu }+1) U_{\mu }^{\dagger}(x)\psi _{x}
\}
+
\overline{\psi}_{x}\psi_{x}
)
\label{eq:wilsonFERMION}
\end{equation}
The tree level relation between $\kappa $ and the quark mass $m$ is $\kappa $ =
$\frac{1}{2(4+m)}$. 

There is a lot of effort in the lattice gauge community on designing
new fermion actions for light quarks (with masses lighter than the
strange quark mass). 
There has been a long standing concern about fermion doubling on the
lattice~\cite{Gupta:1997nd}.  There are a number of pragmatic
solutions to the doubling problem.  For example the action
in equation~\ref{eq:wilsonFERMION} breaks chiral symmetry with an $O(a)$ term.
Chiral symmetry will be restored as the continuum limit ($a
\rightarrow 0$) is taken, but lack of chiral symmetry at finite
lattice spacing causes problems, such as the difficulty of 
reaching light
quark masses.

Our understanding of chiral symmetry on the lattice
has increased by the rediscovery of the 
Ginsparg-Wilson relation~\cite{Ginsparg:1982bj}:
\begin{equation}
M \gamma_5 + \gamma_5 M = a M \gamma_5 M
\label{eq:GWeqn}
\end{equation}
where $M$ is the fermion operator in 
equation~\ref{eq:Faction} at zero mass.

Lattice fermion operators that obey the Ginsparg-Wilson relation
(equation~\ref{eq:GWeqn}) have a form of 
lattice chiral symmetry~\cite{Luscher:1998pq}.
Explicit solutions, such as overlap-Dirac~\cite{Neuberger:2001nb} 
or perfect 
actions~\cite{Hasenfratz:1998jp}, to
equation~\ref{eq:GWeqn} are known.  Actions that obey the
Ginsparg-Wilson relation are increasingly being used for quenched QCD
calculations~\cite{Hernandez:2001yd}.  
This class of actions have not been used for heavy
quark calculations (see~\cite{Liu:2002qu} for some speculations).
Domain Wall actions, that can loosely be thought of as being
approximate solutions to the Ginsparg-Wilson relation are being used
in calculations~\cite{Noaki:2001un,Blum:2001xb} of the matrix elements
for the $\epsilon'/\epsilon$.  Unfortunately, solutions of the
Ginsparg-Wilson relation are too expensive computationally to be used
for unquenched calculations~\cite{Jansen:2001fn}.

A more pragmatic development in the design of light fermion actions is
the development of 
improved staggered fermion actions~\cite{Orginos:1998ue,Bernard:1999xx}.  
This class of action
is being used for unquenched lattice QCD calculations with very light
quarks (see table~\ref{tb:dynamicalPARAMS}) by the MILC collaboration.
The improved staggered quark formalism is quite ugly compared to 
actions that are solutions of the Ginsparg-Wilson relation.
It is not understood why calculations using improved staggered quarks 
are much faster~\cite{Gottlieb:2001cf}
 than calculations using
Wilson fermions~\cite{Lippert:2002jm}.

\begin{figure}[b]
\begin{center}
\includegraphics[width=.8\textwidth]{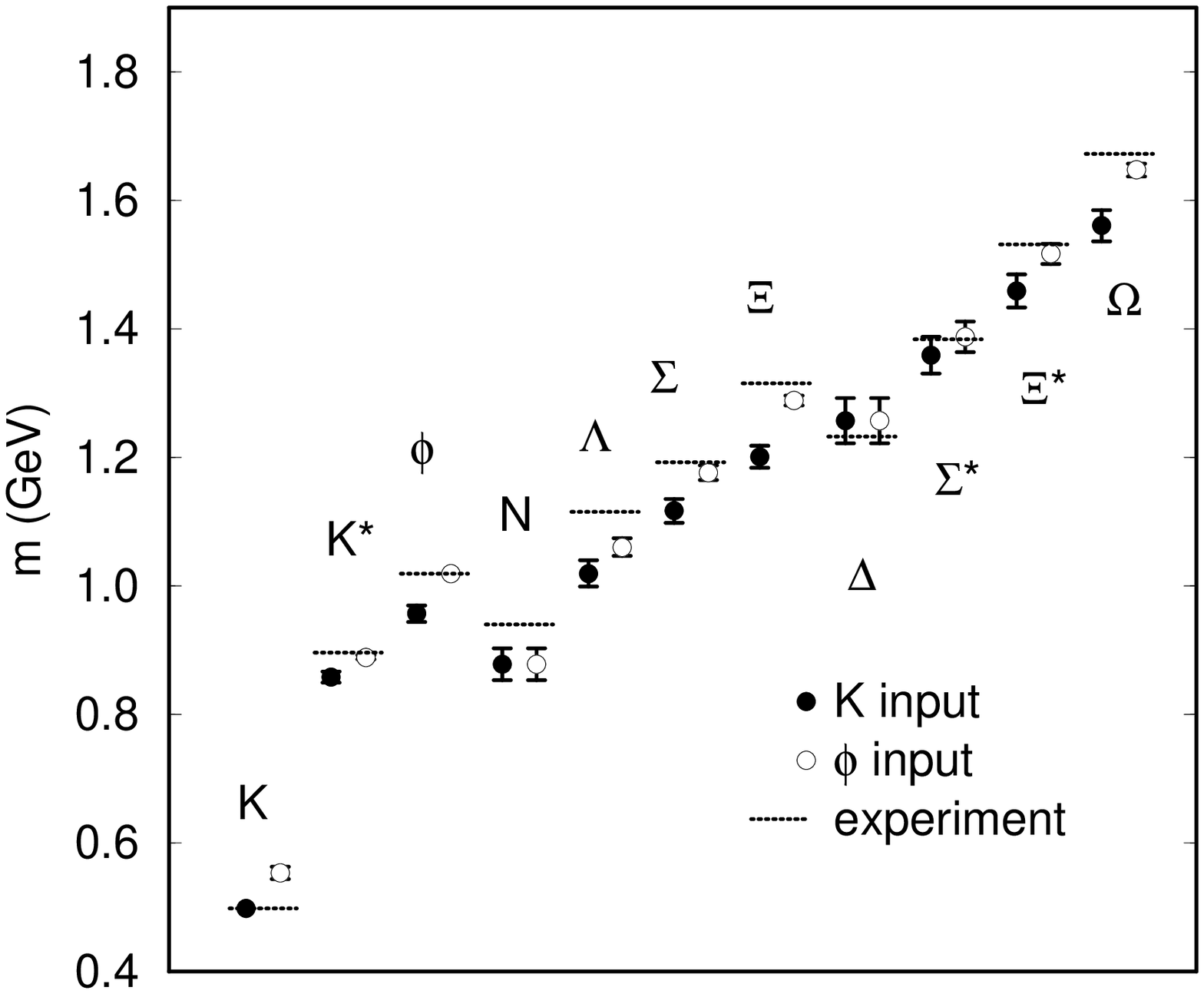}
\end{center}
\caption[]{Spectrum of light hadrons from CP-PACS~\cite{Aoki:1999yr}}
\label{fg:cpPACSlightSPECTRUM}
\end{figure}

The largest systematic study of light hadron spectroscopy
in quenched QCD has been carried out by the 
CP-PACS collaboration~\cite{Aoki:1999yr}.
CP-PACS controlled the systematic errors
by using $a \approx$ 0.1 - 0.05 fm, $m_\pi/m_\rho \approx 0.75$ - 0.4,
and box sizes greater than 3 fm. A summary of CP-PACS's results
is in figure~\ref{fg:cpPACSlightSPECTRUM}
CP-PACS~\cite{Aoki:1999yr} 
summarize their calculation of the light hadron 
spectrum in quenched QCD by the masses
showed a deviation from
experiment of less than 11\%.

Although an agreement between experiment and the results of quenched
QCD at the 11\% level might seem impressive, many heavy-light
matrix elements need to computed to an accuracy of under
5\% to have an impact on tests of the CKM matrix.

There are now indications of the effects of the sea quarks.
For example, the CP-PACS collaboration~\cite{AliKhan:2000mw} have 
used $n_f =2$ lattice QCD to calculate the strange quark mass.
CP-PACS's results are in table~\ref{tab:cppacsstrange}. 
The results show a sizable reduction in the mass of the 
strange quark between quenched and two flavour QCD.
See the review by Lubicz~\cite{Lubicz:2000ch} 
for a discussion of the results of CP-PACS
in comparison to those from other groups.
\begin{table}[tb]
\begin{center}

  \begin{tabular}{|c|c|c|} \hline
$n_f$  & input    &  $m_s$ MeV \\ \hline
2      &  $\phi$  &  $90^{+5}_{-11}$   \\
2      &  $K$     &  $88^{+4}_{-6}$   \\  \hline
0      &  $\phi$  & $132^{+4}_{-6}$   \\
0      &  $K$     & $110^{+3}_{-4}$   \\  \hline
  \end{tabular}
\end{center}
  \caption{
Mass of the strange quark from CP-PACS~\cite{AliKhan:2000mw}
in the $\overline{MS}$ scheme at a scale of 2 GeV.
}
\label{tab:cppacsstrange}
\end{table}
The  results from CP-PACS 
for the strange quark mass need to be checked by other
calculations with lighter quarks and finer lattice spacings,

\section{The different ways of treating heavy quarks on the lattice}

In principle all the above formalism can be used to do
calculations that include charm and bottom quarks.
Unfortunately, in practice there is a restriction that the 
quark mass should be much less than the lattice spacing. 
\begin{equation}
a M_Q << 1 
\end{equation}
As the lattice spacings accessible to current calculations are
$\frac{1}{a} \sim$ 2 Gev (see table~\ref{tb:dynamicalPARAMS}) and 4
GeV~\cite{Becirevic:2002jg} for unquenched and quenched QCD
respectively. Hence calculations using traditional techniques will
just about work for the charm mass ($m_{c}\sim $1.3 GeV), but will not
work for the bottom quark ($m_{b}\sim $5 GeV).  It is computationally
expensive (see equation~\ref{eq:cost}) to reduce the lattice spacing,
so that a $b$ quark will be resolved by the lattice.
There are a variety of special techniques for including the bottom
quark in lattice QCD calculations, all of them are based on heavy
quark effective field theory.

I do not discuss the method of using the potentials measured in lattice
QCD with Schr\"{o}dinger's equation to compute the mass spectrum of
heavy-heavy mesons.  This subject is reviewed by Bali~\cite{Bali:2000gf}.
The potential based approach is not applicable
to computing matrix elements, so is not useful for checks of the 
CKM matrix.

Various subsets of the lattice QCD community have strong opinions on
the right (and wrong) approach to including heavy quarks in lattice
calculations.  Obtaining consistent results from calculations that use
different heavy quark actions is a good check on the systematic
errors.  In the next sections, I describe some of the more popular
techniques used for heavy quarks on the lattice. I report on the
results from this class of methods in section~\ref{se:hyperFINE}.

\subsection{The improvement view} \label{se:improvement}

There are concerns that the results from lattice calculations have
large lattice spacing errors because $aM_{Q}\sim 1$ even for charm
quarks. The Wilson action in equation~\ref{eq:wilsonFERMION}
has $O(a)$ lattice spacing errors. If the $O(a)$ term is removed then
perhaps larger quark masses could be used in the lattice QCD
calculations.

A standard technique from numerical analysis is to use derivatives
that are 
closer approximations to the continuum derivatives.  
For example the lattice derivative in 
equation~\ref{eq:better} 
should be more accurate with a larger lattice spacing than
derivative in equation~\ref{eq:OK}.
\begin{equation}
\frac{f(x+a)-f(x-a)}{2a} = 
\frac{df}{dx}+
  O(a^{2})
\label{eq:OK}
 \end{equation}
\begin{equation}
\frac{4}{3}
\{
  \frac{f(x+ a)-f(x- a)}{2a } 
- \frac{f(x+2a)-f(x-2a)}{16a}
\}
= \frac{df}{dx}+ O(a^{4})
\label{eq:better}
\end{equation}
However in a
quantum field theory there are additional complications, such as
operators mixing under renormalization.

There is a formalism due to
Symanzik~\cite{Symanzik:1983dc,Symanzik:1983gh} called improvement
where new terms are added to the lattice action that cancel $O(a)$
terms in a way that is consistent with quantum field theory.
The ``simplest'' improvement~\cite{Sheikholeslami:1985ij}
 to the Wilson action is to add
the clover term~\ref{eq:cloverTERM} to remove tree level lattice
spacing errors:
\begin{equation}
S_{f}^{clover}= S_{f}^{W} + c_{SW} \frac{i a \kappa }{2}\sum _{x}
(\overline{\psi _{x}}\sigma _{\nu \mu }F_{\nu \mu }\psi _{x})
\label{eq:cloverTERM}
\end{equation}
where $F_{\nu\mu}$  is the lattice field strength tensor.

If the $c_{SW}$ coefficient is computed in perturbation theory is
used then the errors are $O(ag^{4})$. The 
ALPHA collaboration~\cite{Luscher:1997ug} 
have computed $c_{SW}$to all orders in $g^{2}$
using a numerical technique. The result for $c_{SW}$ from ALPHA
is:
\begin{equation}
c_{SW}= \frac{1-0.656g^{2}-0.152g^{4}-0.054g^{6}}{1-0.922g^{2}}
\end{equation}
for $0 < g <1$, where $g$ is the coupling.
The estimate of $c_{SW}$, by ALPHA collaboration, agrees with the one
loop perturbation theory for $g<1/2$.

Some groups tried to use the results from lattice QCD calculations
with quark masses around charm with the 
scaling laws from heavy quark effective field theory to compute
matrix elements for the $b$ quark.
An example tried by the UKQCD
collaboration~\cite{Bowler:2000xw} was to extrapolate the $f_{B}$
decay constant from masses around charm, where the clover action can
be legitimately used, to the bottom mass, using a functional form
based on HQET~\cite{Manohar:2000dt}.
\begin{equation}
\Phi (M_{P}) \equiv (\frac{\alpha (M_{P})}{\alpha (M_{B})})^{2/\beta _{0}}f_{P}
\sqrt{M_{P}} 
= \gamma_P 
(1  
+ \frac{\delta_P}{M_P}  
+ \frac{\eta_P}{M_P^2}  
+ \dots )
\end{equation}
The review by Bernard~\cite{Bernard:2000ki}
describes the potential problems with this approach.
An extrapolation in mass from 2 GeV to around the bottom
quark mass at 5 GeV is problematic. Note 
that UKQCD~\cite{Bowler:2000xw}
did address some of Bernard's criticism~\cite{Bernard:2000ki}

Computers are fast enough to directly include quark masses close to
the bottom mass in quenched calculations.  However, this approach will
not work for unquenched calculations for some time.

\subsection{The static limit of QCD}

It would clearly be better to interpolate in the heavy quark
mass rather than use extrapolations.
The static theory of QCD can be used to compute the properties
mesons with light ante-quark  and static (infinitely heavy) quarks.
A combined analysis of data from static-light and heavy-light
calculations can be used to interpolate to the $b$ quark mass.
This
was the approach taken by the MILC collaboration~\cite{Bernard:1998xi}
in one of the largest calculations of the $f_B$ decay constant.

The lattice static theory of 
Eichten and Hill~\cite{Eichten:1990zv}
\begin{equation}
S_{static} = ia^{3}\sum _{x} b^{\dagger }(x)(b(x)
- U_{0}(x-\widehat{t})b(x-\widehat{t}))
\end{equation}
has been used for $B$ meson physics.

One of the reasons that static quarks have not been included in many
calculations is that it can be difficult to extract masses and
amplitudes using equation~\ref{eq:FITmodel} because the signal to
noise ratio is poor.
However there are numerical techniques~\cite{Michael:1998sg}
that are better, but not in wide spread use in matrix 
element determinations.

To extract matrix elements from static-light calculations requires the
static-light operators to be matched to QCD.  The ALPHA collaboration
have started~\cite{Sommer:2002en} a program to compute the matching
and renormalization factors numerically. The one loop matching factors
are
available.

\subsection{Nonrelativistic QCD}

It would be better to actually do lattice calculations 
at the physical
bottom
or charm quark masses, rather than extrapolate or interpolate to the
physical points. The formalism called nonrelativistic QCD(NRQCD)
allows this~\cite{Thacker:1991bm,Lepage:1992tx}.
NRQCD is an effective field approximation
to QCD for heavy quarks. The operators in the NRQCD Lagrangian
are ordered by the velocity $v$.

A low order Lagrangian for NRQCD is
\begin{equation}
{\cal L}_{0} = \overline{Q}(\Delta_{t}-
\sum_{i=1}^3 \frac{\Delta_{i} \Delta_{-i}}{2M_{Q}a}-
c_{NR}
\frac{\sigma
  .B}{2M_{Q}a}+....)Q
\label{eq:NRQCDL}
\end{equation}
where $\Delta_{\mu}$ are the covariant derivatives on the lattice
and $B$ is the chrmomagnetic field.

The NRQCD formalism works both for both heavy-light (B) and
heavy-heavy systems ($\Upsilon $). 
Estimates from potential models~\cite{Lepage:1993xa}
 suggest that the $v^2 \sim$ 0.1
in $\Upsilon$ and $v^2 \sim$ 0.3 in charmonium. In 
section~\ref{se:hyperFINE}, I review the 
evidence that shows that the NRQCD is not as convergent
in charmonium as the naive power counting arguments suggest.


The main theoretical disadvantage of NRQCD is that the continuum limit
can not be taken because of the $\frac{1}{M_Q a}$ terms in the Lagrangian.
In practice improvement techniques can be used. 

The coefficients, such as $c_{NR}$ in equation~\ref{eq:NRQCDL}, in the
Lagrangian are fixed by matching to QCD.
The matching calculations involve lattice perturbation
theory that is harder than continuum perturbation theory 
because the Feynman rules are more complicated~\cite{Morningstar:1994qe}.

A physically motivated (but not rigorous) way of improving the
convergence of lattice perturbation theory is to use tadpole
improvement~\cite{Lepage:1993xa}. Tadpole perturbation theory can be
used to produce ``reasonable'' tree level estimates for coefficients
in the Lagrangian.

To do perturbation theory the gauge links are expanded in terms of 
the gauge potential:
\begin{equation}
U_{\mu }(x) = e^{iagA_{\mu }(x)} \rightarrow 
1 + i a g A_{\mu }(x) + \dots
\label{eq:naive}
\end{equation}
Equation~\ref{eq:naive} suggests that the $\langle U\rangle \sim 1$,
however this is not seen in numerical lattice calculations.
Also the complicated vacuum structure of QCD would make $\langle
U\rangle \sim 1$
unlikely.
Lepage
and Mackenzie~\cite{Lepage:1993xa} suggest: 
\begin{equation}
U_{\mu }(x) = e^{iagA_{\mu }(x)} \rightarrow 
u_{0}(1 + iagA_{\mu }(x))
\end{equation}
where $u_{0}$ is the ``mean gauge link''. 
Unfortunately there is no unique way of defining
the mean gauge link. The expectation value of the mean
link is zero because it is not gauge invariant.
Some estimates are based on
taking the quartic root of the plaquette
\begin{equation}
u_{0,P} = \langle \frac{1}{3}ReTr(U_{plaq})\rangle ^{1/4}
\end{equation}
or computing the mean link in Landau gauge.

There are projects~\cite{Trottier:1998bn}
under way that attempt to estimate the
coefficients such as $c_{NR}$ to order $\alpha^2$.
The basic idea~\cite{Trottier:2001vj} 
is to try to use weak coupling
numerical lattice QCD calculations to obtain
information on perturbative quantities.

A similar approach to NRQCD is taken by the Fermilab
group~\cite{El-Khadra:1997mp,Kronfeld:2000ck}, 
except that they match to a
relativistic fermion action, essentially the clover action with mass
dependent coefficients.

\subsection{Anisotropic lattices.}

A technique that has been used for heavy 
quarks~\cite{Klassen:1998fh,Chen:2000ej,Okamoto:2001jb}
is to use a lattice spacing that is smaller in the time direction than
in the space direction to circumvent the restriction $a m_Q << 1$.
A finer lattice spacing is used in the time 
direction such that $a_{t}m_{q}\leq 1$
but a larger lattice spacing $a_s$ is used the spatial direction 
to keep the cost down and stop any problems with finite size effects.
This approach assumes that the discretization error is
only weakly dependent on $a_s m_q$.

The anisotropic clover 
operator~\cite{Okamoto:2001jb} is
\begin{equation}
M=m_{0}+\nu_{0}\widehat{W_{0}}
\gamma _{0}+\frac{\nu }{\xi _{0}}
\sum_{i}\widehat{W_{i}}\gamma _{i}
+ \frac{i}{2}(w_{0}\sum _{x,i}
\sigma_{0i}\widehat{F_{0i}}(x)+
\frac{w} {\xi _{0}}\sum _{x,i<j} \sigma_{ij}F_{ij}(x)) 
\end{equation}
The clover action (equation~\ref{eq:wilsonFERMION} plus~\ref{eq:cloverTERM}) 
is reproduced by the conditions:
$\xi_0 = 1$, $\nu$ = $\nu_0$, and $w$ = $w_0$.

The parameters: $w_{0}$, $w_{i}$, $\nu_{0}$, and $\nu$ need to be correctly
chosen. For example Klassen~\cite{Klassen:1998fh} proposed to tune 
$\nu$, by computing the pseudoscalar meson mass at nonzero
momentum, and to choose the value of $\nu$ that gave 
$c(\underline{p}) = 1$.
\begin{equation}
E(\underline{p})^2 = E(0)^2 + c(\underline{p})^2  \underline{p}^2
\end{equation}
Klassen's~\cite{Klassen:1998fh}  original motivation
for using this class of action was that is was potentially easier
to tune the unknown parameters using the techniques developed
by the ALPHA collaboration~\cite{Luscher:1997ug}
than for the Fermilab heavy quark
action~\cite{El-Khadra:1997mp}.

The pure gauge action is also 
modified~\cite{Okamoto:2001jb}
\begin{equation}
S_{g} = \beta (\frac{1}{\xi _{0}}\sum _{x,s>s'}[1-P_{ss'}(x)]+\xi
_{0}\sum _{x,s}[1-P_{ss}(x)]) 
\end{equation}
where $P_{ss}$ are purely spatial plaquettes and $P_{ss'}$
are plaquettes in space and time.
The renormalized anisotropy $\xi _{0}=a_{s}/a_{t}$ (
ratio of lattice spacings in time and space)
can be measured by comparing the lattice
potential in space and time~\cite{Alford:2000an}.

A practical problem in lattice calculations is that the 
signal in equation~\ref{eq:FITmodel} is lost in the
noise for large times. The smaller lattice spacing in the 
time direction from anisotropic lattices means that the 
region in lattice units, where there is a signal, is longer,
thus it is easier to fit equation~\ref{eq:FITmodel} to
the data. 
Collins at al.~\cite{Collins:2001pe} used this feature
of anisotropic lattices to get improved signals for form
factors.
Although the fit region in lattice units is longer,
the actual fit region in physical units may be smaller,
this may cause problems. 


\subsection{The hyperfine splitting in charmonium} \label{se:hyperFINE}

\begin{figure}[b]
\begin{center}
\includegraphics[width=.8\textwidth]{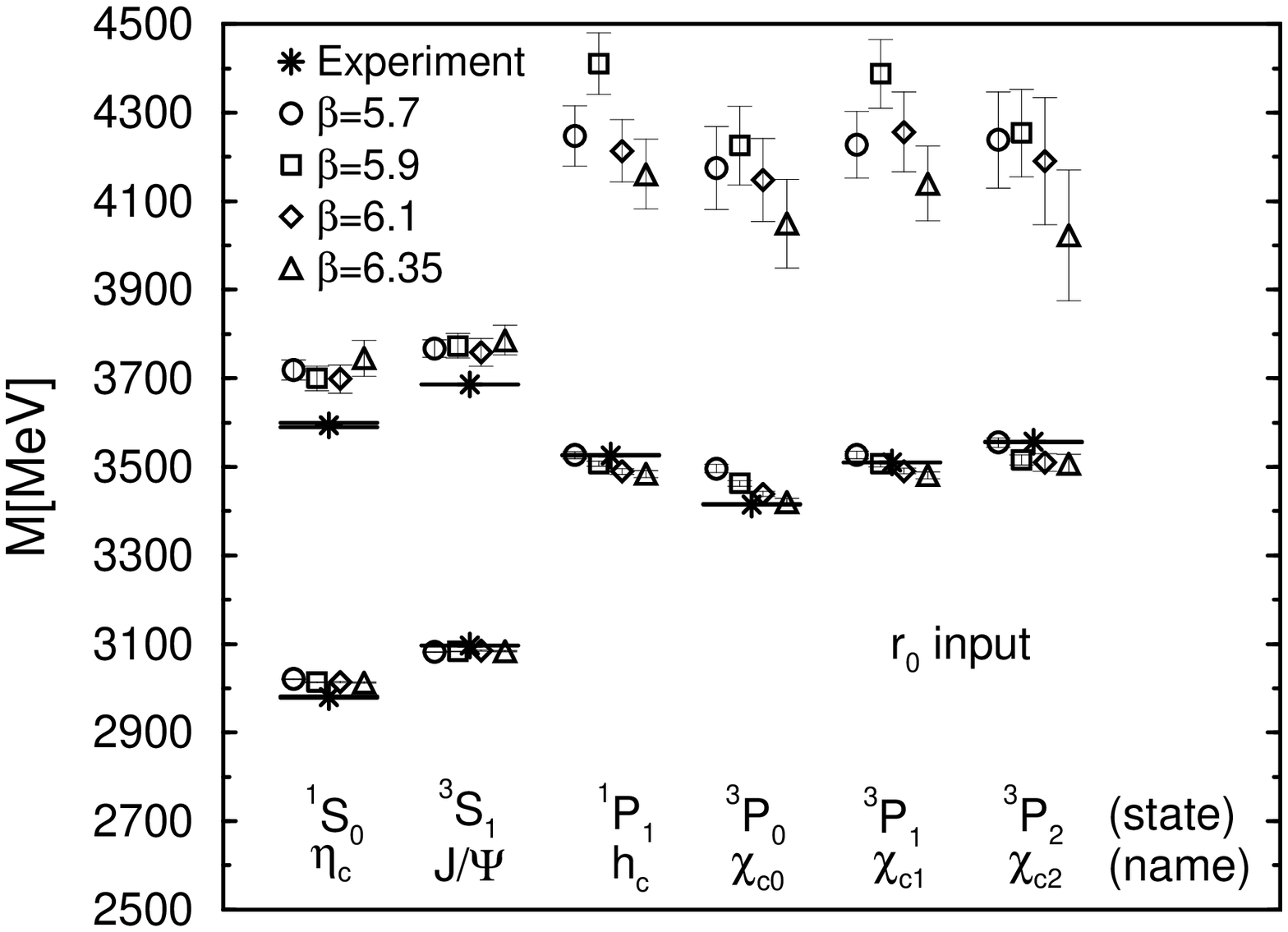}
\end{center}
\caption[]{Spectrum of charmonium from CP-PACS~\cite{Okamoto:2001jb}}
\label{fg:cpPACScharm}
\end{figure}

It is obviously important to test the methods used to solve lattice
QCD by comparing the results against experiment. This validation
procedure ensures that the various errors in the calculations are
under control. Figure~\ref{fg:cpPACScharm} shows the charmonium
spectrum from lattice QCD calculations by the CP-PACS
collaboration~\cite{Okamoto:2001jb}. The overall agreement with
experiment is quite good.

A particularly good test of lattice QCD techniques is to compute the
mass splitting between the $J/\psi$ and $\eta_c$. 
In the section I will use $\Delta m_H$ to denote the 
mass difference between $J/\psi$ and $\eta_c$. 
The masses of these
two meson can usually be computed with the smallest statistical error
bars. Also, as these masses are independent of light valence quarks,
this splitting does not depend on a chiral extrapolation of the
valence quarks.
Hein et al.~\cite{Hein:2000qu} discuss the
various systematic errors in the mass splittings
between other heavy hadrons.


I will start by discussing the results from quenched QCD.
I will use $\Delta m_H$ as the mass splitting between the 
$J/\psi$ and $\eta_c$.
The experimental value for the mass splitting between the $J/\psi$ and
$\eta_c$ is 116 MeV.  
It was therefore disappointing that some of the
lattice QCD calculations in the early 90's were:
$\Delta m_H$ = 
28(2) MeV and 52(4) MeV from the Wilson action and tree level 
clover action respectively~\cite{Allton:1992zy} ($a^{-1}$ = 2.73 GeV
from the string tension). 

Using the clover action with the a value of the clover coefficient
$c_{SW}$ (see equation~\ref{eq:cloverTERM}) motivated by tadpole
perturbation theory, the Fermilab group~\cite{El-Khadra:1993ir}
obtained $\Delta m_H$ = $93 \pm 10$ MeV.

The hyperfine splitting is sensitive $c_{SW}$ at nonzero lattice
spacing, but the hyperfine splitting should be independent of the
$c_{SW}$ as the continuum limit is taken, because the clover term is
an irrelevant operator.

Recently the QCD-TARO collaboration~\cite{Choe:2001yg} have studied
the charmonium spectrum using the clover action at a smaller lattice
spacing ($a^{-1} \sim$ 5 GeV) than previously used. Using the clover
action QCD-TARO collaboration~\cite{Choe:2001yg} obtained $\Delta m_H$
= $99 \pm 7$ MeV, $\Delta M$ = $87 \pm 2$ MeV, for quadratic and
linear extrapolations in the lattice spacing to the continuum.

The NRQCD collaboration calculated 
$\Delta m_H$  to be 96(2) MeV~\cite{Davies:1995db} using an 
NRQCD action that included terms of $O(m_c v^{4})$. The lattice
spacing was
$a^{-1}$ = 1.23(4) GeV. Using power accounting arguments 
the size of the next order in the NRQCD expansion was estimated to
be 30 to 40 MeV. Unfortunately, when 
Trottier~\cite{Trottier:1997ce} included the 
$O(m_c v^6)$ relativistic corrections, he obtained
$\Delta m_H = $  55(5) MeV. 
The hyperfine splitting is sensitive to the $c_{NR}$ coefficient
(see equation~\ref{eq:NRQCDL}). Further work by 
Shakespeare and Trottier~\cite{Shakespeare:1998dt} showed
that the hyperfine splitting was sensitive to the 
tadpole prescription used to estimate $c_{NR}$, so the 
final word on the utility of NRQCD for charm quarks may have to
wait for $c_{NR}$ to be computed beyond one loop. 
The caveat is that the current estimates for 
 $c_{NR}$  seem to produce good agreement with the 
hyperfine splittings in the baryon sector~\cite{Mathur:2002ce}.

Although NRQCD is clearly not the technique to use to compute the
hyperfine splitting in charmonium, NRQCD may be valid for hadrons with
charm quarks, such as the mass splittings between the S-wave states
and the speculated $1^{-+}$ state~\cite{Manke:1998qc} and D
mesons~\cite{Hein:2000qu}.

There was a preliminary attempt~\cite{Orginos:1998fh}
 to use 
an action motivated by  renormalisation group arguments
(perfect action~\cite{Hasenfratz:1998ft})
to compute the charmonium spectrum. This class of action should
produce results with reduced lattice spacing dependence.
Unfortunately the action used in~\cite{Orginos:1998fh} did not
produce a result for the hyperfine splitting closer to experiment
than any other approach.

Klassen~\cite{Klassen:1998fh} first used anisotropic lattices to study the 
charmonium system. Using spatial lattice spacings in the range 
0.17 to 0.3 fm (the scale was set using $r_0 = 0.5 fm$)
and two anisotropies 2 and 3, Klassen obtained a hyperfine
splitting of just over 90 MeV in the charmonium system.
Chen~\cite{Chen:2000ej} 
obtained $\Delta m_H = $ 71.8(2.0) MeV
with anisotropy $\xi _{0}=a_{s}/a_{t}$ =2.

The definitive study of the anisotropic lattice technique
for charmonium spectroscopy was carried out by
CP-PACS~\cite{Okamoto:2001jb}. They fixed the anisotropy at
3 and used spatial lattice spacings between 0.07 and 2 fm
(finer then both Chen~\cite{Chen:2000ej} and 
Klassen~\cite{Klassen:1998fh}). The results from
CP-PACS~\cite{Okamoto:2001jb}
were: $\Delta m_H = $ = 73(4) MeV using $r_0$ to set the scale 
and 85(8) MeV using the $P$-$S$ splitting for the lattice
spacing.
I have combined the different 
errors using quadrature.
CP-PACS concluded that $a_s m_q < 1 $ is still required for a 
reliable continuum extrapolation.

A qualitative explanation for the low value of the 
hyperfine splitting in charmonium from quenched QCD
 was given by 
El-khadra~\cite{El-Khadra:1993ir} 
using potential model ideas.
In El-khadra's model the 
Richardson potential~\cite{Richardson:1979bt}
is used
\begin{equation}
V(q^2) = C_F \frac{4 \pi} {\beta_0^{n_f}}
\frac{1}{q^2 \log (1 + q^2/\lambda^2)}
\end{equation}
with
\begin{equation}
\beta_{0}^{n_f} = 11 - 2 n_f /3 
\end{equation}
to solve for the wave function of the charm quark.
The wave function depends on the number of flavours
$n_f$. El-khadra obtained the result
\begin{equation}
\frac{ \Psi^0(0) } {\Psi^3(0)} = 0.86
\end{equation}
In this model the hyperfine splitting is related to the 
wave function and coupling ($\alpha_s$) as 
\begin{equation}
\Delta m_H
\sim \frac{\alpha_s(m_c) }{m_c^2}
\mid \Psi (0) \mid^2
\end{equation}
Including the suppression of the coupling in the quenched
theory,  El-khadra estimated 
\begin{equation}
\Delta m_H^{quenched} \sim 70 MeV.
\end{equation}
There have been a number of unquenched lattice 
QCD 
calculations~\cite{Glassner:1996xi,Allton:1998gi,Allton:2001sk,Bernard:2000gd}
that have seen evidence for the $n_f$ dependence of the 
heavy quark 
potential
at small distances.
The MILC collaboration~\cite{Bernard:2000gd}
have systematically studied the wave functions
from the measured heavy quark potential from quenched and unquenched
calculations.

There has not been much work on the charmonium spectrum from
unquenched lattice QCD calculations.
El-Khadra et al.~\cite{El-Khadra:2000zs} 
did look at the charmonium spectrum on 
(unimproved) staggered gauge configurations from the MILC
collaboration. 
No significant increase in the hyperfine
splitting was reported.
The $m_{\pi}/m_{\rho}$ was 0.6 and the lattice spacing was 
$a^{-1} \sim $ 0.99(4) GeV.

Stewart and Koniuk~\cite{Stewart:2000ev} studied the charmonium
spectrum using NRQCD on unquenched (unimproved ) staggered 
gauge configurations ($m_{\pi}/m_{\rho} \sim$ 0.45 
and $a \sim$ 0.16 fm). Any signal for the effect of unquenching
was hidden beneath the other systematic uncertainties in 
using NRQCD for charmonium.

Although the potential model argument of El-Khadra for the 
effect of quenching on the hyperfine splitting gives some
insight, it does not explain the  sea quark mass dependence
of the splitting. Grinstein and Rothstein~\cite{Grinstein:1996gm}
have developed a formalism
based on Chiral Lagrangian for the dependence of quarkonium mass
splittings between 1P-1S and 2S-1S on the sea quark mass.
Up to chiral logs they predict for the splitting $\delta m$
\begin{equation}
\delta m \sim A + B m_{\pi}^2
\label{eq:chiralSPLIT}
\end{equation}
where $A$ and $B$ are unknown parameters and 
$m_{\pi}$ is the mass of the pion made out of light
sea quarks.

In my opinion a decade's worth of lattice QCD calculations
of the $J/\psi$ -$\eta_c$ mass splitting can be summarized as 
waiting for unquenched gauge configurations with light sea quark masses.
Recently, progress has been
made in the Upsilon system using the gauge configurations generated by
the MILC collaboration. The preliminary 
work by Gray et al~\cite{Gray:2002vk} found that the
correct ratio was produced for the (P-S)/(2S-1S) mass splittings in
Upsilon.  It will be interesting to see the charmonium spectrum on
these lattices, particularly if relations such as~\ref{eq:chiralSPLIT}
can be tested and used.

\begin{figure}[b]
\begin{center}
\includegraphics[width=.8\textwidth]{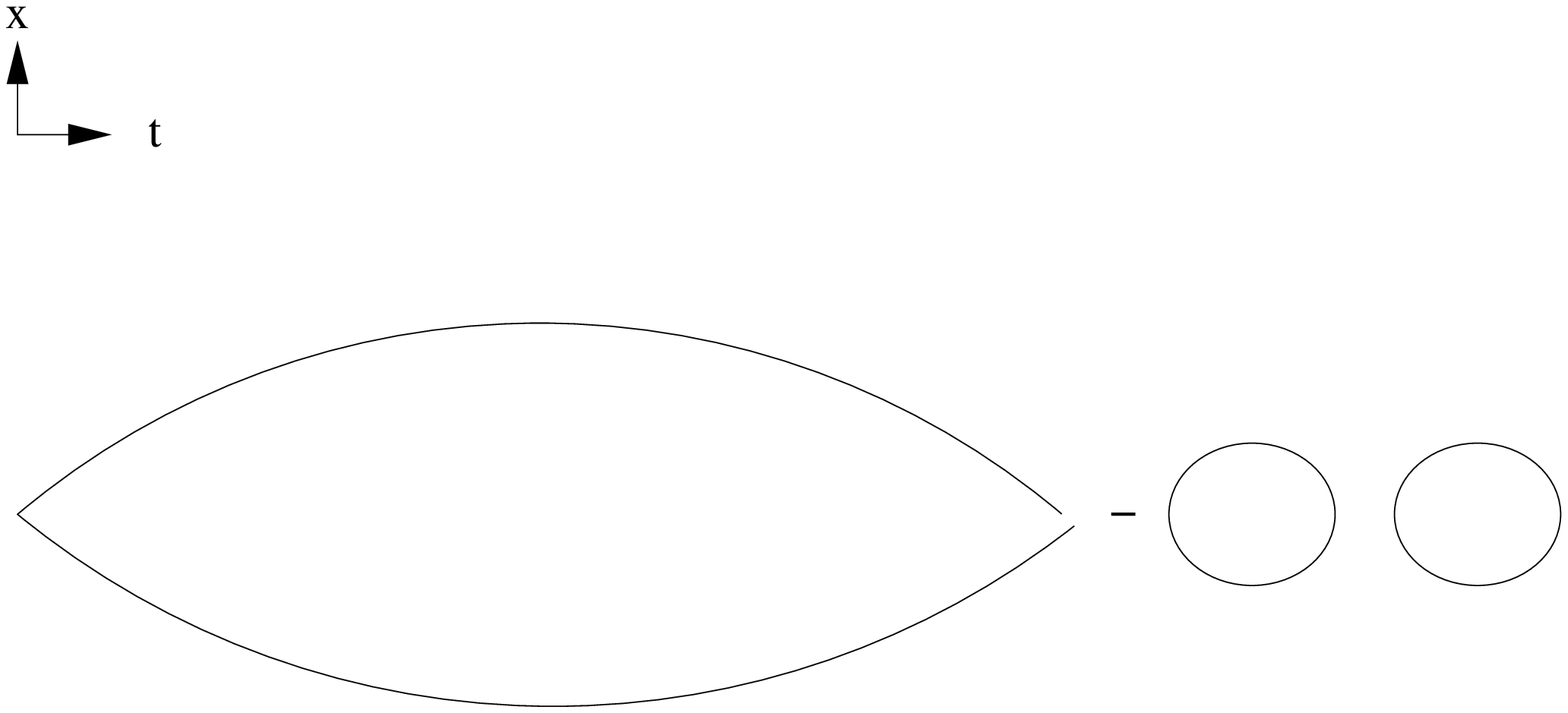}
\end{center}
\caption[]{Two point correlator}
\label{fg:singletTWOpoint}
\end{figure}

There is another possible reason that the hyperfine mass splitting 
between the $J/\psi$ and $\eta_c$ is smaller than experiment
in current simulations. All lattice calculations have
computed the non-singlet correlator 
(see figure~\ref{fg:twopoint}). However, 
charmonium interpolating operators are actually singlet, so
the Wick  contractions contain bubble diagrams (see 
figure~\ref{fg:singletTWOpoint}).
The bubble diagrams are OZI suppressed so should be small.
However, this argument will fail if there is additional nonperturbative
physics. For light mesons~\cite{McNeile:2001cr}, 
it has been found that the effects of 
the bubbles can be large for the pseudoscalar and scalar
mesons where the additional physics is the anomaly and 
the $0^{++}$ glueball, but not for other channels.

Morningstar and Peardon~\cite{Morningstar:1999rf} 
have computed the glueball spectrum
in quenched QCD. They obtained masses of 2590(40)(130) MeV
and 3640(60)(180) MeV for the ground and first excited states
of the $0^{-+}$ glueball respectively. Morningstar and Peardon computed the 
mass of the $1^{--}$ glueball to be 3850(50)(190) MeV. So it is not
inconceivable that the $\eta_c$ mass (2980 MeV) is effected more
by glueball states than the $J/\psi$ state.
The above comments are speculations and can be checked by 
explicit lattice calculations. As the effect of the bubble diagrams
is almost certainly less than 50 MeV, hence this will be a very hard 
mass splitting to estimate.

The mass spectrum of heavy-light mesons introduces the additional
complication of the light valence quark. Lattice QCD calculations can
be done with quark masses around charm, but for computational reasons
the light quarks have masses that are typically greater than half the
strange quark mass. The lattice data is extrapolated in the light
quark mass to the physical points using a fit model based on chiral
symmetry.

In table~\ref{tab:hyperfine} I have collected some results for 
the $D^\star-D$ mass splitting.
\begin{table}[tb]
\begin{center}
\begin{tabular}{|c|c|c|} \hline
Group & Method &  $M_{D^{\star}}-M_{D}$ MeV \\  \hline
Boyle~\cite{Boyle:1997aq} & clover &  124(5)(15) \\
Boyle~\cite{Boyle:1998rk} & $\beta$=6.0 tadpole clover &  106(8) \\
Hein et al.~\cite{Hein:2000qu} & NRQCD  $\beta = 5.7$ & 
$103^{+3+22}_{-0-0}(3)(6)(5)$ \\
UKQCD~\cite{Bowler:2000xw} & NP clover  $\beta = 6.2$ & 
$130^{+6+15}_{-6-35}$
\\ \hline 
PDG~\cite{Groom:2000in} & Experiment  & 140.64(10) 
\\ \hline 
  \end{tabular}
\end{center}
  \caption{
Collection of hyperfine splittings between the
$D$ and \protect{$D^{\star}$}.
}
\label{tab:hyperfine}
\end{table}
Currently there is a lot of effort in the lattice gauge theory
community to study the chiral extrapolations of quantities with the  
light quark mass.
The Adelaide 
group~\cite{Leinweber:1999ig,Leinweber:2001ac}
have developed various phenomenological forms
for the light quark mass dependence of hadron masses loosely motivated by
effective field theories. The fit models have had some empirical
success with extrapolating the 
masses of the rho and nucleon~\cite{Leinweber:1999ig,Leinweber:2001ac}, with
the caveat that lattice spacing errors were not taken into account.
Similar techniques were applied to the 
$B^\star-B$
and $D^\star-D$ 
mass splittings 
by Guo and Thomas~\cite{Guo:2001ph}. No improvement
with the agreement between experiment and the lattice data was
seen.

\section{Case study: calculating the charm mass from lattice QCD}

The general steps involved in many lattice QCD calculations are fairly
similar.  To explain the component parts of a lattice QCD calculation,
I will explain the use of lattice QCD data to extract the charm quark
mass from experimental data. Full details of the calculation of the
charm quark mass can be found in~\cite{Becirevic:2001yh,Rolf:2002gu}. I
will not discuss the approach~\cite{Kronfeld:1998zc,Juge:2001dj} to
computing the charm quark mass based on Fermilab's heavy quark
action~\cite{El-Khadra:1997mp}) and the pole mass.  There is a useful
review by El-Khadra and Luke~\cite{El-Khadra:2002wp} on computing the
mass of bottom quark.

The error for the mass of the charm quark quoted in the particle data
table is 8\% - an unbelievably large error for a basic parameter of
the standard model that was discovered in 1974.  The experimental mass
of the $D$ meson is $1869.3 \pm 0.5$, hence the error on the charm
quark mass is predominantly from theory.  There are many places in particle
physics where a more accurate value of the charm mass would be useful.
Some models of quark matrices predict relationships between quark
masses and CKM matrix elements. For example Fritzsch and
Xing~\cite{Fritzsch:2001nv} predict
\begin{equation}
\frac{\mid V_{ub}\mid }{\mid V_{cb}\mid } = \sqrt{\frac{m_{u}}{m_{c}}}
\end{equation}
To test such relations we need to accurately determine all the
component parameters of the standard model.

The starting point, I shall take is the masses of the heavy-light
mesons as a function of lattice parameters. The hadron
masses  come from a fit of the 
correlator in equation~\ref{eq:FITmodel}.
For example, table~\ref{tb:UKQCDheavyLight}
contains the masses of a heavy-light meson in lattice units 
$\beta =6.2$ from UKQCD~\cite{Bowler:2000xw}.
The $\kappa$ value is defined in the action in 
equation~\ref{eq:wilsonFERMION}.
After this point no supercomputers are required,
just a nonlinear $\chi ^{2}$ fitting program, physical insight and theoretical
physics.
\begin{table}[tb]
\begin{center}
\begin{tabular}{|c|c|c|c|}
\hline 
$\kappa _{H}$&
$\kappa _{L}$&
$aM_{P}$&
$aM_{V}$\\
\hline
\hline 
0.1200&
0.1346&
$0.841_{-1}^{+1}$&
$0.871_{-2}^{+2}$\\
\hline 
0.1200&
0.1351&
$0.823_{-1}^{+2}$&
$0.856_{-2}^{+2}$\\
\hline 
0.1233&
0.1346&
$0.739_{-1}^{+1}$&
$0.775_{-2}^{+2}$\\
\hline
\end{tabular}
\end{center}
\caption{Heavy-light meson mass 
from UKQCD~\cite{Bowler:2000xw}
 as a function of 
$\kappa$ value. $M_{PS}$ and $M_{V}$ are the pseudoscalar and 
vector meson masses}
\label{tb:UKQCDheavyLight}
\end{table}

To start the journey from the lattice to the real world,
the lattice parameters need to be converted to more 
physical parameters. The first job is to convert from the 
kappa value into the quark mass. There are a number of different
expressions for the  quark mass in terms of the lattice parameters.

One definition of the  quark mass is based on the 
vector ward identity.
\begin{equation}
m_V = \frac{1}{2} ( \frac{1}{\kappa} - \frac{1}{\kappa_{crit}} )
\label{eq:mdefn}
\end{equation}
The $m_V$ quark mass suffers from an additive renormalisation for 
Wilson like fermions. The value of $\frac{1}{\kappa_{crit}}$
is obtained by the value of $\kappa$ that gives zero pion mass.
To remove $O(a)$ corrections the improvement 
formalism requires that 
\begin{equation}
\hat{m} = m ( 1 + b_m m a )
\end{equation}
The perturbative value 
of $b_m$ is $-\frac{1}{2} - 0.096 g^2$~\cite{Sint:1997jx}.

The quark mass can also be defined in terms of the PCAC relation.
\begin{equation}
m_{AW} = \frac{ \langle \partial_4 A_4(t) P(0) \rangle } 
             { 2 \langle P(t) P(0) \rangle  }
\label{eq:PcAcMass}
\end{equation}
There are also O(a) corrections to equation~\ref{eq:PcAcMass}, 
see~\cite{Rolf:2002gu}
for details.

In principle the masses $m_V$ and $m_{AW}$ should agree, however
at finite lattice spacing, where the calculations are actually
done they disagree. For quark masses below strange, it has been
shown that the two definitions agree as the lattice spacing is
taken to zero (see~\cite{Gupta:1997sa} for a review).

The masses must be converted from lattice units into
MeV.
As explained in section~\ref{se:intro}, one quantity must be sacrificed
to find the lattice spacing. For example at $\beta$ = 6.2 in quenched QCD,
UKQCD find $a^{-1}$ = $2.66^{+7}_{-7}$, 
$2.91_{+1}^{-1}$, $2.54_{+4}^{-9}$, GeV from 
$f_{\pi}$, $r_0$, and $m_{\rho}$  respectively.
The spread in different choices should reduce as the continuum
limit is taken in an unquenched lattice QCD calculation.
If there are different choices of lattice spacing, this is
usually included in the systematic error.

Now we have a table of data heavy-light meson masses in GeV versus
the lattice quark masses in GeV. The meson masses must be 
interpolated and extrapolated to the physical meson masses.  The
theory behind the extrapolations is an effective Lagrangian 
for mesons with heavy quark  and chiral symmetry~\cite{Manohar:2000dt}.

The value of $\kappa$ corresponding to the strange quark mass is
usually determined from light quark spectroscopy by interpolating to
the mass of kaon or phi meson.
Becirevic et al.~\cite{Becirevic:2001yh} investigated
using three different fit models to extrapolate the meson masses in the
heavy quark mass.
\begin{eqnarray}
M_{HL}^1(m_Q) & = &  A_{1} + B_{1} (m_Q) +  C_{1} m_Q^2 \\
M_{HL}^2(m_Q) & = &  A_{2} + B_{2} (\frac{1}{m_Q}) 
+  C_{2} (\frac{1}{m_Q^2}) \label{eq:hqetMODEL} \\
M_{HL}^3(m_Q) & = &  A_{3} + B_{3} (\frac{1}{m_Q}) 
+  C_{3} m_Q 
\end{eqnarray}

The $1/M_Q$ terms 
are motivated by heavy quark symmetry~\cite{Manohar:2000dt}.
Kronfeld and Simone~\cite{Kronfeld:2000gk} 
have used the fit model in equation~\ref{eq:hqetMODEL} 
to estimate $\lambda_1$ and $\lambda_2$ parameters of HQET.

\subsection{Quark mass renormalization factors}

In the last section I showed how to find the charm quark mass
in the lattice scheme. However, normally quark masses are
used in application in the $\overline{MS}$ scheme. Also, a
consistent scheme is also required to compare the results 
from different calculations. 

The quark mass in the lattice scheme
($m_L(a) $) is matched to the $\overline{MS}$ 
scheme $m_{\overline{MS} }(\mu)$.
\begin{equation}
m_{\overline{MS} }(\mu)  = Z_{m}(a \mu) m_L(a) 
\end{equation}
where $Z_{m}(a \mu) $ is the matching factor.

The matching factor has been computed in perturbation theory
to one loop order.
\begin{equation} 
Z_{m}(a \mu) = 1 - \frac{\alpha(\mu)}{4 \pi}
(  8 \ln (\mu a ) -  C_M   ) 
\end{equation}
The value of $C_M$ depends on the 
fermion action.
For the clover action,
$C_M = 25.8$~\cite{Gockeler:1998fn}.

As usual with one loop calculations, there is an ambiguity as to
what scale ($\mu$) to evaluate the matching at
The ``best guess scale'' for the $\mu$ (called $q\star$) can
in principle be computed using the formalism described by Lepage
and Mackenzie~\cite{Lepage:1993xa}. Most people include the effect
of varying $\mu$ in some range from $1/a$ to $\pi/a$ in 
the systematic errors.

The accuracy of the quark mass determination would improve if the
matching could be done to higher order than one loop.  The Feynman
rules on the lattice are more complicated than in the continuum, hence
calculations beyond one loop are very hard. Some groups are starting
to try to automate lattice perturbation theory~\cite{Drummond:2002yg}.

The general framework of lattice QCD allows other approaches to
computing matching factors without using Feynman diagrams 
on the lattice.
Sint~\cite{Sint:2000vc} and Sommer~\cite{Sommer:2002en} review some of
the ways that matching factors are computed on the lattice.  For
example the $\alpha^3$ term of the residual mass (important for the
extraction of the bottom quark mass) of static theory in quenched QCD
was computed using a numerical technique~\cite{DiRenzo:2000nd}.

A general technique~\cite{Martinelli:1995ty} for matching between the
lattice and $\overline{MS}$ schemes has been developed by the
Rome and Southampton groups. The basic idea is to use the quark
propagator calculated in lattice QCD to do the lattice part of the
matching. The gauge has to be fixed in this approach. Usually Landau
gauge is chosen.

\subsection{Evolving the quark mass to a reference scale}

The matching procedure produces the charm mass at the matching scale.
To compare different mass determinations, the quark mass has to be
evolved to a standard reference scale, essentially the same as that
used by the particle data table. The reference scale chosen for the charm
quark is the charm quark mass itself.

The running quark mass equation is used to evolve the quark mass
to the standard reference scale of the charm quark
mass. 
\begin{equation}
\mu ^{2}\frac{d}{d\mu ^{2}}m^{n_{f}}(\mu ) = m^{n_{f}}(\mu )\gamma _{m}^{n_{f}}(\alpha _{s}^{n_{f}})
= -\sum _{i\geq 0}\gamma _{m,i}^{n_{f}}(\frac{\alpha
  _{s}^{n_{f}}(\mu )}{\pi })^{i+1}
\end{equation}
The coefficients $\gamma _{m,i}^{n_{f}}$ are 
known to four loop order.
The required equations are conveniently packaged in the RunDec
Mathematica package~\cite{Chetyrkin:2000yt}.
The ALPHA collaboration have a method to do the 
evolution of the quark mass numerically~\cite{Capitani:1998mq,Sommer:2002en}.
The method is starting to be used for 
unquenched QCD~\cite{Chetyrkin:2000yt}.

\subsection{Comparison of the results}

I have outlined the basic ideas behind a lattice QCD calculation of
the charm quark mass.  In table~\ref{mcmcEveryOneElse} I collect the
state-of-the art results for the charm quark mass from (quenched) lattice QCD.
These should be compared with the result quoted in the particle
data table of  $m_c^{\overline{MS}}(m_c)$ between 
1.0 to 1.4 GeV. 

\begin{table}[tb]
\begin{center}
  \begin{tabular}{|c|c|} \hline
Group &  $m_c^{\overline{MS}}(m_c) GeV$
 \\  \hline
Becirevic  et al.~\cite{Becirevic:2001yh}&  1.26(4)(12)  \\
Rolf and Sint~\cite{Rolf:2002gu} &   1.301(34)  \\
Juge~\cite{Juge:2001dj} & 1.27(5) \\
Kronfeld~\cite{Kronfeld:1998zc} &   1.33(8) \\ \hline
  \end{tabular}
\end{center}
  \caption{
The charm quark mass from
quenched lattice QCD.
}
\label{mcmcEveryOneElse}
\end{table}

\section{The $f_{B}$ decay constant}

A crucial quantity for tests of the
CKM matrix formalism is equation~\ref{eq:BBmixing}
\begin{equation}
\frac{\Delta m_s}{\Delta m_d}  = 
\mid \frac{ V_{ts} }{  V_{td}   } \mid^2
\frac{m_{B_s}}{m_{B_d}} 
\xi^2
\label{eq:BBmixing}
\end{equation}
The value of $\Delta m_d$ has been measured experimentally, while
$\Delta m_s$ is expected to be measured at run II of the
Tevatron~\cite{Anikeev:2001rk}. There are already useful
experimental limits on $\Delta m_s$.
The hard part is extracting the ratios of 
QCD matrix elements in
\begin{equation}
\xi^2 = \frac{f_{Bs}^2B_{s}}{f_{B}^2 B}
\end{equation}
The quantity $\xi$ can not be extracted from experiment
and is non-perturbative.

The $f_{B}$ decay is the QCD matrix element for the semi-leptonic
decay of the B meson. It is analogous to the pion decay constant.
It is claimed that $f_{B}$ will never be measured experimentally,
hence it must be computed from QCD. The $B$ (bag) factors are also 
QCD matrix elements that have been computed from lattice QCD.

The computation of $f_{B}$ shares many features to the calculation
of the charm quark mass. The same data from supercomputers could
be used for both the $f_{B}$ and charm quark mass calculation.
$f_B$ is extracted from the matrix element
\begin{equation}
\langle 0\mid A_{0}\mid Qq,p=0\rangle  = -if_{Qq}M_{Qq}
\end{equation}
This matrix element is simply related to the amplitudes 
($c_{ij}$) in equation~\ref{eq:FITmodel}.
The main additional complication over the charm mass calculation is
the extrapolation of the decay constant to the
bottom mass.

Sinead Ryan reviewed the latest results for the $f_{B}$ decay constant
at the lattice 2001 conference~\cite{Ryan:2001ej}. Ryan's world average
of the lattice data for heavy light decay constants is in
table~\ref{tb:DECAYav}.
\begin{table}[tb]
\begin{center}\begin{tabular}{|c|c|}
\hline 
$N_{f}$  &  Decay constant  \\ \hline
0        &  $f_{B}$ = 173(23) MeV   \\
2        &  $f_{B}$ = 198(30) MeV   \\
0        &  $f_{Ds}$ = 230(14) MeV  \\
2        &  $f_{Ds}$ = 250(30) MeV  \\
\hline
\end{tabular}\end{center}
\caption{Summary~\cite{Ryan:2001ej} of the results for heavy-light decay constants
from quenched $N_{f}=0$ and ($N_{f}=2$) unquenched lattice QCD.}
\label{tb:DECAYav}
\end{table}

The lattice methods can be checked by computing the $f_{D_s}$ decay
constant.  The current experimental result for $f_{Ds}$ = 280 (40)(19)
MeV~\cite{Groom:2000in}. The CLEO-c experiment plans to reduce the
experimental errors on $f_{Ds}$ to the few percent
level~\cite{Shipsey:2002ye} to test lattice QCD. 
The actual comparison
between theory and experiment will be 
ratios of matrix
elements for leptonic and semi-leptonic decays, so that the 
test is independent of CKM matrix elements.

In her review article Ryan~\cite{Ryan:2001ej} 
quoted the errors on $\xi$ from lattice QCD as
\begin{equation}
\xi = 1.15(6)_{-0}^{+7}
\label{eq:xiWORLDav}
\end{equation}

The first error in equation~\ref{eq:xiWORLDav} is the statistical and
systematic errors from quenched QCD.  The asymmetric errors are from
unquenching.  It is instructive to compare the errors on $\xi$ with
the experimental errors on $\Delta m_d$ = $0.503 \pm 0.006 ps$.
Although $\Delta m_s$ has not yet been measured, it is expected to be
measured to a few percent accuracy at the Tevatron.  The final errors
on $\mid\frac{V_{ts} }{ V_{td} }\mid^2 $ will be limited by the
theoretical errors on $\xi$.

Unfortunately, during the last year the errors on $\xi$ have gone up
again, based on some observations by the JLQCD
collaboration~\cite{Yamada:2001xp}.  Kronfeld and
Ryan~\cite{Kronfeld:2002ab} have suggested that a more realistic value
of $\xi$ is $1.32 \pm 0.10$ rather than the estimate in
equation~\ref{eq:xiWORLDav}.

The key problem is that the light quarks in the current unquenched
lattice QCD calculations are not so light.  Lattice QCD calculations
are typically done at unphysically large mass parameters.  Physical
results are obtained by extrapolating the results using effective
field theories.

The effective field theory for heavy-light systems contains the light
particles: $\pi$, $K$, and $\eta$, and a pseudoscalar and vector
heavy-light state~\cite{Manohar:2000dt}.  The Lagrangian is written so
that it is invariant under heavy quark symmetry and $SU(3)_L \times
SU(3)_R$ symmetry.  This Lagrangian is for static quarks.  The
Lagrangian can be used to calculate masses of hadrons and decay
constants in terms of the couplings in the Lagrangian.

The most important coupling at tree level is the $g_{\pi}$
coupling that describes the $D^{\star} \rightarrow D + \pi$ 
decay (suitably extrapolated to the heavy quark limit).
Table~\ref{tb:gpiRESULTS} contains some estimates of $g_{\pi}$
from experiment and lattice QCD.

\begin{table}[tb]
\begin{center}\begin{tabular}{|c|c|c|}
\hline 
Group & method & $g_{\pi}$ \\ \hline
UKQCD~\cite{deDivitiis:1998kj}   & Lattice QCD & $0.42(4)(8)$ \\
Abada et al.~\cite{Abada:2002xe} & Lattice QCD & $0.69(18)$ \\ \hline
CLEO~\cite{Anastassov:2001cw}    & Experiment $D \star D\pi$ & $0.59 \pm 0.01 \pm 0.07$   \\
\hline
\end{tabular}\end{center}
\caption{Summary of some results for $g_{\pi}$
}
\label{tb:gpiRESULTS}
\end{table}

The first loop correction to the decay constant has the form
\begin{equation}
\sqrt{m_B} f_B = \Phi( 1 + \Delta f_q)
\end{equation}
where $\Phi$ is the quantity with zero light quark
mass and $\Delta f_q$ represents the deviation from 
the chiral limit due to the finite size of the light quark
mass.

The problems with the chiral extrapolations of $\xi$ are due 
to the ratio of
the decay constants, so consider:
\begin{eqnarray}
\xi_f -1 & \equiv & \frac{f_{B_s}}{f_B} - 1 \\
         & = & \delta f_s - \delta f_d \\
         & = & (m_K^2  - m_\pi^2 ) f_2(\mu) + C 
\label{eq:fBMODEL}
\end{eqnarray}
where $f_2(\mu)$ is a low energy constant of the 
effective field theory. 
The form of $C$ is 
\begin{eqnarray}
C = \frac{1+ 3g_{\pi}^2}{(4 \pi)^2  }
( 
  \frac{1}{2} m_K^2 \ln ( \frac{m_K^2}{\mu^2})  
+ \frac{1}{4} m_\eta^2 \ln ( \frac{m_\eta^2}{\mu^2}  )
- \frac{3}{4} m_\pi^2 \ln ( \frac{m_\pi^2}{\mu^2}  
 )
\end{eqnarray}
The equivalent expression for the bag parameters $B$ has the 
coefficient $1 - 3 g_{\pi}^2 $ in front of the chiral
logs, from the values of $g_{\pi}$ in table~\ref{tb:gpiRESULTS}, the
$m_\pi^2 \ln ( \frac{m_\pi^2}{\mu^2} )$ term has a negligible effect.

Until recently most lattice QCD calculations 
extrapolated $\xi$ with the $C$ function set to zero.
For example, the MILC collaboration used linear
and quadratic chiral extrapolations into their 
fits for their original results~\cite{Bernard:1998xi}.
The JLQCD collaboration tried to fit 
equation~\ref{eq:fBMODEL} to their unquenched data.

Kronfeld and Ryan~\cite{Kronfeld:2002ab} noted that once the $g_{\pi}$
in known, then the chiral log term in $C$ is known.  Hence, they used
the lattice data that is essentially consistent with linear quark mass
dependence
and
add the log term by hand. The problem with this type of approach is
that it assumes that the current lattice data is in the regime where
there are no higher order corrections to 
equation~\ref{eq:fBMODEL}.  The
definitive answer for the value of $\xi$ will come from unquenched
calculations with light quarks that explicitly see the chiral logs in
$f_B$.

This ``case study'' demonstrates the importance of the parameters of
the dynamical quarks to the computation of heavy-light matrix
elements, particularly the masses of the sea quarks.  This study also
demonstrates that the use of quenched QCD to compute
heavy-light matrix elements is coming to an end.  The chiral structure
of matrix elements in quenched QCD can be very different to that in
unquenched QCD.

\subsection{Computation of form factors from lattice QCD}

One of the best ways to extract the $\mid V_{cb} \mid$ 
CKM matrix element from experiment is to use the
$B \rightarrow  D^{\star} l \nu_l$ semi-leptonic 
decays~\cite{Neubert:1994mb,Manohar:2000dt}.
The differential decay rate~\cite{Artuso:2002zh},
based on HQET~\ref{eq:isguWiseEXPT}
 is
\begin{equation}
\frac{d \Gamma}{d w} (B \rightarrow D^{\star}l\nu_l)
=
\frac{G_F^2 \mid V_{cb} \mid^2} {48 \pi^2}
{\cal K}(w) {\cal F}(w)^2
\label{eq:isguWiseEXPT}
\end{equation}
where ${\cal} K(w)$ is a known  phase space factor
and ${\cal F}(w)^2$ a  form factor. The value of 
$w$ is the dot product of the velocities of the two heavy-light
mesons. The expression in equation
is based on heavy quark effective field theory.

As the masses of the $b$ and $c$ quarks go to infinity the 
normalization point of the Isgur-Wise function is obtained
${\cal F}(1) =1 $.
The form factor at zero recoil is broken into the following
\begin{equation}
{\cal F}(1) = \eta_{QED} \eta_A ( 1 + \delta_{1/m_Q^2} + ..)
\label{eq:beyondLUKE}
\end{equation}
where $\eta_{QED}$ is a perturbative QED factor and
$\eta_A$ is the perturbative matching factor between QCD and
HQET.
The $\delta_{1/m_Q^2}$ term represents the breaking of
heavy quark effective field theory. A term of the form 
$\frac{1}{m_Q}$
is forbidden by Luke'e theorem~\cite{Luke:1990eg}.

The value of $\mid V_{cb} \mid $ in the particle
data table~\cite{Artuso:2002zh} 
from $B \rightarrow D^{\star} l \nu$ decays
is:
\begin{equation}
\mid V_{cb} \mid_{exclusive} = ( 42.1 \pm 1.1_{expt} \pm 1.9_{theory}) 
x 10^{-3}
\end{equation}
The theoretical error is dominated by the theoretical uncertainty in 
$\delta_{1/m_Q^2}$. In the past $\delta_{1/m_Q^2}$ has been computed using
sum rules and quark models. 
Without a systematic way of improving the results, 5\%
will be a lower limit on the accuracy of this CKM matrix element.

\begin{figure}[b]
\begin{center}
\includegraphics[width=.8\textwidth]{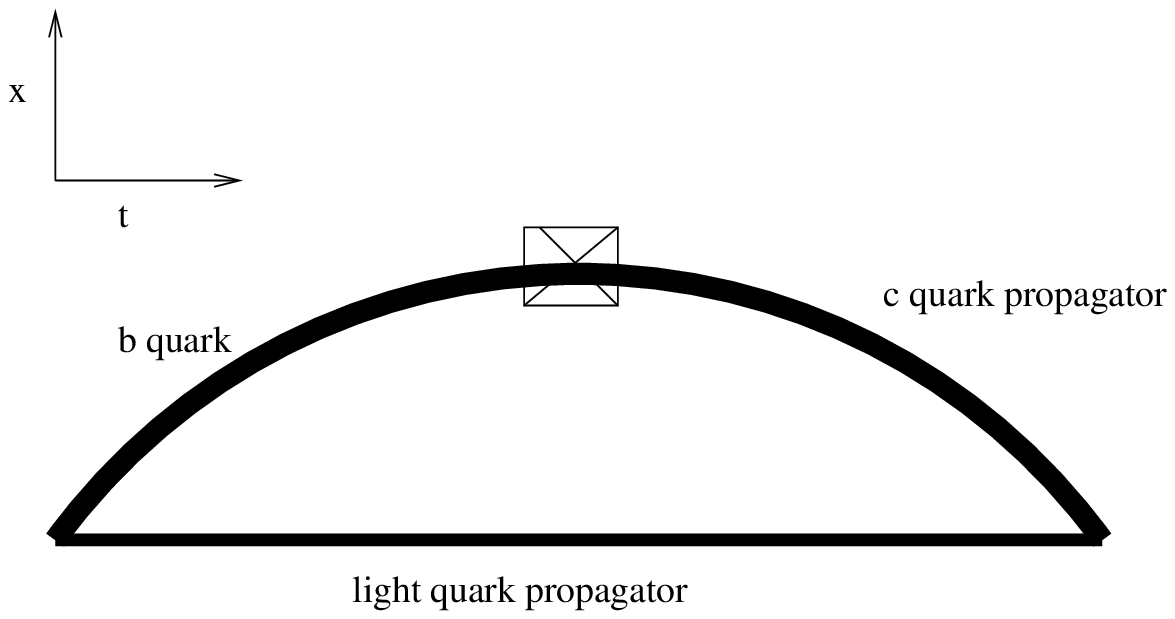}
\end{center}
\caption[]{Space time diagram of three point function}
\label{fg:3pt}
\end{figure}

In figure~\ref{fg:3pt} I show a space-time diagram
that is used to calculate form factors for semi-leptonic
using the path integral.
The initial lattice 
studies ~\cite{Bernard:1993ey,Booth:1994zb,Bowler:1995bp,Bowler:2002zh}.
mapped out the 
dependence of the $B \rightarrow D$ form factors 
on $w$.
The experimental data is taken at nonzero recoil 
$w <> 1 $. However, the extrapolation to $w=1$~\cite{Briere:2002ew}
is either done using a simple ansatz or 
using the results of a 
dispersion relation~\cite{Caprini:1998mu,Boyd:1997kz}.

The Fermilab group~\cite{Hashimoto:2001nb} have concentrated on
estimating $\delta_{1/m_Q^2}$ in equation~\ref{eq:beyondLUKE}.  They
compute the matrix element in figure~\ref{fg:3pt} with all the mesons
at rest. By taking clever combinations of matrix elements they can get
a precise estimate of the form factor at zero recoil.  
The mass dependence of the form factor can be mapped out by 
varying the masses
of the heavy quarks. The matching to the continuum is quite
involved.
The final
results~\cite{Hashimoto:2001nb} have errors with the same order of
magnitude as other approaches. Their error includes a 10\%
estimate for unquenching.

There are other ways that lattice QCD can contribute to the extraction
of the $\mid V_{cb} \mid$ CKM matrix element from experimental data.
There have been calculations of the semi-leptonic decays of the
$\Lambda_b$ baryon~\cite{Bowler:1998ej} from lattice QCD.  There have
been two calculation of the mass of the $B_c$
meson~\cite{Jones:1998ubw,Shanahan:1999mv}.  Jones and
Woloshyn~\cite{Jones:1998ubw} computed the decay constant for the
leptonic decay of $B_c$ to be 420(13) MeV (the error is statistical
only), using NRQCD at a lattice spacing of 0.163(3) fm.
Whether these additional channels can help reduce the 
theoretical uncertainty of $\mid V_{cb} \mid$ to below
the 5\% level is not clear.

\section{Nonleptonic decays}

One of the main goals of the B physics experimental program is to
check the CKM matrix formalism by measuring the CKM matrix elements
many different ways~\cite{Stone:2001jh,Wise:2001ii}.
For example, the experimental measurements for $B^{+} \rightarrow
\pi^{+}\pi^{0}$ could be used to extract $\sin (2\alpha)$ if the
hadronic uncertainties could be controlled.

The path integral in equation~\ref{eq:PATH} is calculated in Euclidean
space to regulate the oscillations in Minkowski space. This means that
the amplitudes extracted from lattice QCD calculations are always
real.
Recently, there has been some theoretical work
(see~\cite{Ishizuka:2002nm} for a review)
on the non-leptonic decays of the kaons,
motivated by the attempts to compute the hadronic matrix elements
for $\epsilon'/\epsilon$.

There were some early attempts to study the decays $D \rightarrow K
\pi$ on the lattice~\cite{Bernard:1990ii,Abada:1990ns,Bernard:1991dp}.
These type of lattice calculations stopped when the theoretical
problems with making contact with experiment became apparent.  In this
section I briefly describe some of the old work on the $D \rightarrow
K \pi$ decays and provide pointers to the new theoretical
developments.

The correlator required is
\begin{equation}
G(t) = 
\langle 0 \mid 
(\pi K) (t) H_{eff}(0) D(t_K) \mid 0 \rangle
\label{eq:DKpi}
\end{equation}
where $D(t_K)$ is the interpolating field for the $D$ meson
at time $t_K$ and $\pi K$ is the interpolating operator for
the pion and kaon. 
The effective Hamiltonian is 
\begin{equation}
H_{eff} = 
c_{+} (\mu) O^{cont}_{+}  + 
c_{-} (\mu) O^{cont}_{-}
\end{equation}
\begin{equation}
O^{cont}_{\pm} = 
( \overline{s_L}\gamma_{\mu}c_L    )
(\overline{u_l}\gamma_{\mu}d_L) 
 \pm
(\overline{s_L}\gamma_{\mu}d_L)
(\overline{u_L}\gamma_{\mu}c_L)
\end{equation}
where $c_{\pm}(\mu)$ are perturbative coefficients.

\begin{figure}[b]
\begin{center}
\includegraphics[width=.8\textwidth]{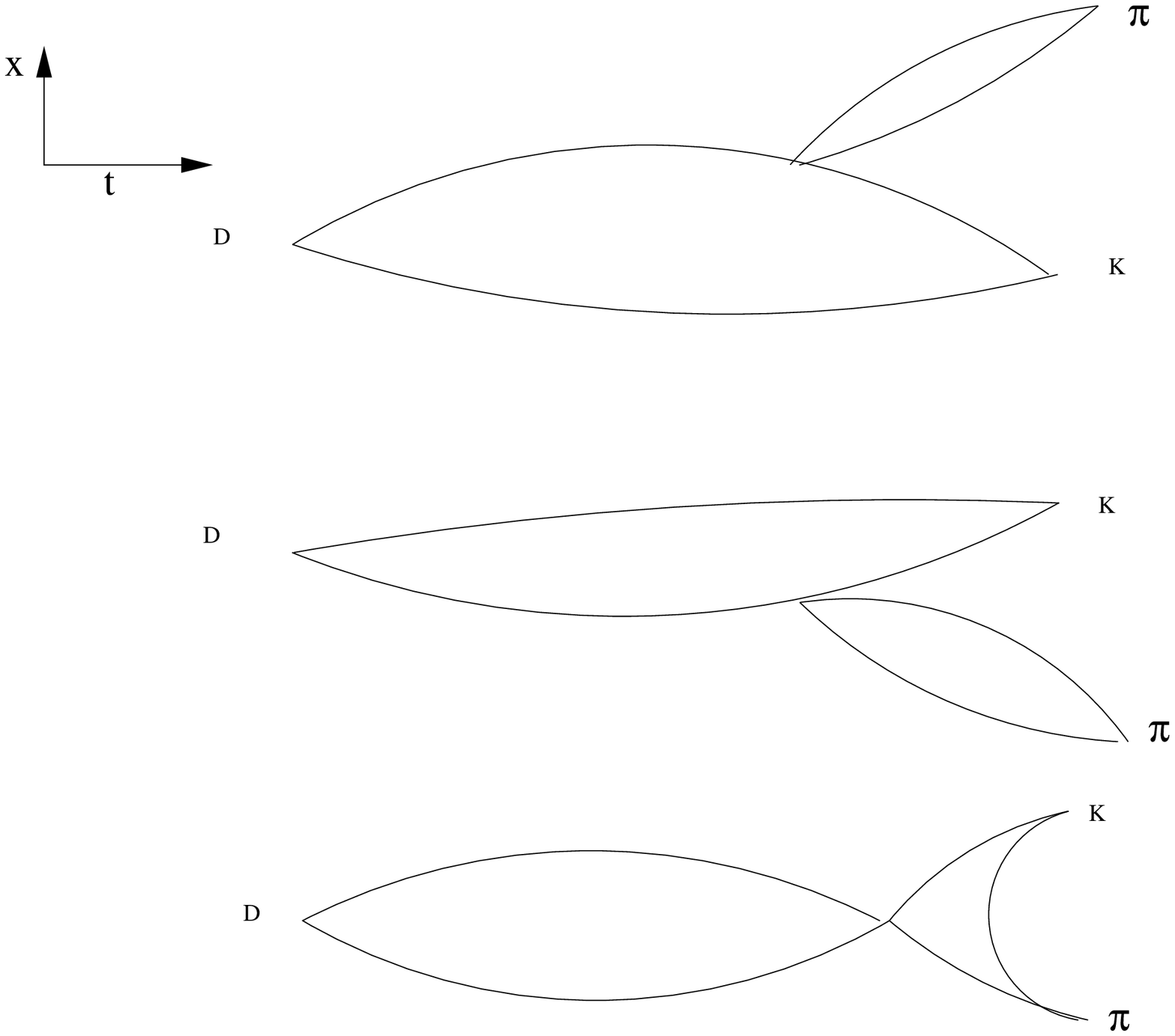}
\end{center}
\caption[]{Wick contractions for equation~\ref{eq:DKpi} }
\label{fg:fourPtWick}
\end{figure}

The diagrams for the Wick contraction of equation~\ref{eq:DKpi} are in 
figure~\ref{fg:fourPtWick}.
Although the diagrams are more complicated to compute than those
for leptonic or semileptonic decays they  can be calculated on a
supercomputer. The problems occur trying to extract the pertinent
amplitudes from $G(t)$ in equation~\ref{eq:DKpi}.

As pointed out by Michael~\cite{Michael:1989mf} and, Maiani and 
Testa~\cite{Maiani:1990ca} there is a complication with creating
a pion and kaon state with definitive momentum. The operator
\begin{equation}
O_{\pi K}(t) =  \pi(\underline{p},t) K(-\underline{p},t)
\label{eq:oper}
\end{equation}
has the same quantum numbers as $\pi K$  with all possible
momentum values. The ground state  of the operator in 
equation~\ref{eq:oper} will be the pion and kaon at rest.
Hence, in the analogue of equation~\ref{eq:FITmodel}
for this graph, the required amplitude will not be the ground state.
It is not easy to fit a multi-exponential model to data,
although it is possible with a basis of 
interpolating operators~\cite{Ishizuka:2002nm}.

Theoretical work, in the context of $K \rightarrow \pi\pi$ decays has
shown how to get matrix elements in infinite volume from matrix
elements computed in finite volumes~\cite{Lellouch:2000pv,Lin:2001ek}.
There have also been proposals~\cite{Michael:1989mf,Ciuchini:1996mq}
to introduce some model independence to extract the complex phases of
the matrix elements.

The methodology for non-leptonic decays 
will be further developed and tested
on $K \rightarrow \pi \pi$ decays, before any 
attempts are made at the decays $D \rightarrow K \pi$,

\section{Conclusions}

The consumers of lattice QCD results need the error bars on current
matrix elements to be reduced below 5\%.  The hardest error to reduce
is from quenching. Improved staggered quarks look like they will be
the first to explore unquenched QCD with light sea quarks.  This
should motivate the champions of other light quark formalisms to speed
up their unquenched calculations.  The techniques that will reduce the
error bars of heavy-light matrix elements will lie outside the domain
of heavy quarks, in areas such as algorithms, improved computer
hardware, and better grant writing.

My own, admittedly biased view, is that quenched QCD calculations are
now of limited use for lattice QCD calculations with heavy quarks.
As every experimentalist I have ever met has held this view,
I am sure it will prevail.

The computation of matrix elements for two body hadronic decays
still looks quite hard. Interesting things seem to be happening for
kaon decays and in theory. It is not clear, whether these developments
will be useful for non-leptonic decays of the $B$ meson.  It would
obviously be a major breakthrough if this problem could be solved,
however Mark Wise's wise~\cite{Wise:2001ii} words, on the career
ending nature of working on nonleptonic decays should be heeded.

The computation of the QCD matrix elements for the heavy flavour
program is a well defined task.  If we can't compute them reliably,
then we will have failed. We will have to admit that we can't
compute anything from QCD outside perturbation theory from first
principles. I hope this doesn't happen.

\section{Acknowledgments}

I thank Chris Michael and  Alex Dougall for discussions.



\begin{thebibliography}{100}

\bibitem{Neubert:1994mb}
M.~Neubert,
\newblock Phys. Rept. {\bf 245}, 259 (1994), hep-ph/9306320,
\newblock 

\bibitem{Manohar:2000dt}
A.~V. Manohar and M.~B. Wise,
\newblock Cambridge Monogr. Part. Phys. Nucl. Phys. Cosmol. {\bf 10}, 1 (2000),
\newblock 

\bibitem{Draper:1998ms}
T.~Draper,
\newblock Nucl. Phys. Proc. Suppl. {\bf 73}, 43 (1999), hep-lat/9810065,
\newblock 

\bibitem{Bernard:2000ki}
C.~W. Bernard,
\newblock Nucl. Phys. Proc. Suppl. {\bf 94}, 159 (2001), hep-lat/0011064,
\newblock 

\bibitem{Ryan:2001ej}
S.~M. Ryan,
\newblock Nucl. Phys. Proc. Suppl. {\bf 106}, 86 (2002), hep-lat/0111010,
\newblock 

\bibitem{Davies:1998ur}
C.~T.~H. Davies {\em et~al.},
\newblock (1998), hep-lat/9801024,
\newblock 

\bibitem{Mueller-Preussker:2002cp}
M.~Mueller-Preussker {\em et~al.},
\newblock (2002), hep-lat/0203004,
\newblock 

\bibitem{Kronfeld:2002pi}
A.~S. Kronfeld,
\newblock (2002), hep-lat/0205021,
\newblock 

\bibitem{Davies:1997hv}
C.~Davies,
\newblock (1997), hep-ph/9710394,
\newblock 

\bibitem{Davies:2002cx}
C.~Davies,
\newblock (2002), hep-ph/0205181,
\newblock 

\bibitem{Flynn:1998ca}
J.~M. Flynn and C.~T. Sachrajda,
\newblock Adv. Ser. Direct. High Energy Phys. {\bf 15}, 402 (1998),
  hep-lat/9710057,
\newblock 

\bibitem{Gupta:1997nd}
R.~Gupta,
\newblock (1997), hep-lat/9807028,
\newblock 

\bibitem{Flynn:2002yu}
J.~Flynn, L.~Lellouch, and G.~Martinelli,
\newblock (2002), hep-lat/0209167,
\newblock 

\bibitem{Negele:1998ev}
J.~W. Negele,
\newblock Nucl. Phys. Proc. Suppl. {\bf 73}, 92 (1999), hep-lat/9810053,
\newblock 

\bibitem{Maynard:2001zd}
UKQCD, C.~M. Maynard,
\newblock Nucl. Phys. Proc. Suppl. {\bf 106}, 388 (2002), hep-lat/0109026,
\newblock 

\bibitem{Lippert:2002jm}
TXL, T.~Lippert,
\newblock Nucl. Phys. Proc. Suppl. {\bf 106}, 193 (2002), hep-lat/0203009,
\newblock 

\bibitem{Gottlieb:2001cf}
S.~Gottlieb,
\newblock Nucl. Phys. Proc. Suppl. {\bf 106}, 189 (2002), hep-lat/0112039,
\newblock 

\bibitem{DiPierro:2001yu}
M.~Di~Pierro,
\newblock Nucl. Phys. Proc. Suppl. {\bf 106}, 1034 (2002), hep-lat/0110116,
\newblock 

\bibitem{McNeile:2000qm}
UKQCD, C.~McNeile,
\newblock (2000), hep-lat/0003009,
\newblock 

\bibitem{Davies:2002mu}
UKQCD, C.~T.~H. Davies, A.~C. Irving, R.~D. Kenway, and C.~M. Maynard,
\newblock (2002), hep-lat/0209121,
\newblock 

\bibitem{Bernard:1999xx}
MILC, C.~W. Bernard {\em et~al.},
\newblock Phys. Rev. {\bf D61}, 111502 (2000), hep-lat/9912018,
\newblock 

\bibitem{AliKhan:2000mw}
CP-PACS, A.~Ali~Khan {\em et~al.},
\newblock Phys. Rev. Lett. {\bf 85}, 4674 (2000), hep-lat/0004010,
\newblock 

\bibitem{Allton:2001sk}
UKQCD, C.~R. Allton {\em et~al.},
\newblock Phys. Rev. {\bf D65}, 054502 (2002), hep-lat/0107021,
\newblock 

\bibitem{Glassner:1996xi}
TXL, U.~Glassner {\em et~al.},
\newblock Phys. Lett. {\bf B383}, 98 (1996), hep-lat/9604014,
\newblock 

\bibitem{Sharpe:1998hh}
S.~R. Sharpe,
\newblock (1998), hep-lat/9811006,
\newblock 

\bibitem{Gray:1990yh}
N.~Gray, D.~J. Broadhurst, W.~Grafe, and K.~Schilcher,
\newblock Z. Phys. {\bf C48}, 673 (1990),
\newblock 

\bibitem{Sommer:1994ce}
R.~Sommer,
\newblock Nucl. Phys. {\bf B411}, 839 (1994), hep-lat/9310022,
\newblock 

\bibitem{Ginsparg:1982bj}
P.~H. Ginsparg and K.~G. Wilson,
\newblock Phys. Rev. {\bf D25}, 2649 (1982),
\newblock 

\bibitem{Luscher:1998pq}
M.~Luscher,
\newblock Phys. Lett. {\bf B428}, 342 (1998), hep-lat/9802011,
\newblock 

\bibitem{Neuberger:2001nb}
H.~Neuberger,
\newblock Ann. Rev. Nucl. Part. Sci. {\bf 51}, 23 (2001), hep-lat/0101006,
\newblock 

\bibitem{Hasenfratz:1998jp}
P.~Hasenfratz,
\newblock Nucl. Phys. {\bf B525}, 401 (1998), hep-lat/9802007,
\newblock 

\bibitem{Hernandez:2001yd}
P.~Hernandez,
\newblock Nucl. Phys. Proc. Suppl. {\bf 106}, 80 (2002), hep-lat/0110218,
\newblock 

\bibitem{Liu:2002qu}
K.-F. Liu,
\newblock (2002), hep-lat/0206002,
\newblock 

\bibitem{Noaki:2001un}
CP-PACS, J.~I. Noaki {\em et~al.},
\newblock (2001), hep-lat/0108013,
\newblock 

\bibitem{Blum:2001xb}
RBC, T.~Blum {\em et~al.},
\newblock (2001), hep-lat/0110075,
\newblock 

\bibitem{Jansen:2001fn}
K.~Jansen,
\newblock Nucl. Phys. Proc. Suppl. {\bf 106}, 191 (2002), hep-lat/0111062,
\newblock 

\bibitem{Orginos:1998ue}
MILC, K.~Orginos and D.~Toussaint,
\newblock Phys. Rev. {\bf D59}, 014501 (1999), hep-lat/9805009,
\newblock 

\bibitem{Aoki:1999yr}
CP-PACS, S.~Aoki {\em et~al.},
\newblock Phys. Rev. Lett. {\bf 84}, 238 (2000), hep-lat/9904012,
\newblock 

\bibitem{Lubicz:2000ch}
V.~Lubicz,
\newblock Nucl. Phys. Proc. Suppl. {\bf 94}, 116 (2001), hep-lat/0012003,
\newblock 

\bibitem{Becirevic:2002jg}
SPQ(CD)R, D.~Becirevic, V.~Lubicz, and C.~Tarantino,
\newblock (2002), hep-lat/0208003,
\newblock 

\bibitem{Bali:2000gf}
G.~S. Bali,
\newblock Phys. Rept. {\bf 343}, 1 (2001), hep-ph/0001312,
\newblock 

\bibitem{Symanzik:1983dc}
K.~Symanzik,
\newblock Nucl. Phys. {\bf B226}, 187 (1983),
\newblock 

\bibitem{Symanzik:1983gh}
K.~Symanzik,
\newblock Nucl. Phys. {\bf B226}, 205 (1983),
\newblock 

\bibitem{Sheikholeslami:1985ij}
B.~Sheikholeslami and R.~Wohlert,
\newblock Nucl. Phys. {\bf B259}, 572 (1985),
\newblock 

\bibitem{Luscher:1997ug}
M.~Luscher, S.~Sint, R.~Sommer, P.~Weisz, and U.~Wolff,
\newblock Nucl. Phys. {\bf B491}, 323 (1997), hep-lat/9609035,
\newblock 

\bibitem{Bowler:2000xw}
UKQCD, K.~C. Bowler {\em et~al.},
\newblock Nucl. Phys. {\bf B619}, 507 (2001), hep-lat/0007020,
\newblock 

\bibitem{Bernard:1998xi}
C.~W. Bernard {\em et~al.},
\newblock Phys. Rev. Lett. {\bf 81}, 4812 (1998), hep-ph/9806412,
\newblock 

\bibitem{Eichten:1990zv}
E.~Eichten and B.~Hill,
\newblock Phys. Lett. {\bf B234}, 511 (1990),
\newblock 

\bibitem{Michael:1998sg}
UKQCD, C.~Michael and J.~Peisa,
\newblock Phys. Rev. {\bf D58}, 034506 (1998), hep-lat/9802015,
\newblock 

\bibitem{Sommer:2002en}
R.~Sommer,
\newblock (2002), hep-lat/0209162,
\newblock 

\bibitem{Thacker:1991bm}
B.~A. Thacker and G.~P. Lepage,
\newblock Phys. Rev. {\bf D43}, 196 (1991),
\newblock 

\bibitem{Lepage:1992tx}
G.~P. Lepage, L.~Magnea, C.~Nakhleh, U.~Magnea, and K.~Hornbostel,
\newblock Phys. Rev. {\bf D46}, 4052 (1992), hep-lat/9205007,
\newblock 

\bibitem{Lepage:1993xa}
G.~P. Lepage and P.~B. Mackenzie,
\newblock Phys. Rev. {\bf D48}, 2250 (1993), hep-lat/9209022,
\newblock 

\bibitem{Morningstar:1994qe}
C.~J. Morningstar,
\newblock Phys. Rev. {\bf D50}, 5902 (1994), hep-lat/9406002,
\newblock 

\bibitem{Trottier:1998bn}
H.~D. Trottier and G.~P. Lepage,
\newblock Nucl. Phys. Proc. Suppl. {\bf 63}, 865 (1998), hep-lat/9710015,
\newblock 

\bibitem{Trottier:2001vj}
H.~D. Trottier, N.~H. Shakespeare, G.~P. Lepage, and P.~B. Mackenzie,
\newblock Phys. Rev. {\bf D65}, 094502 (2002), hep-lat/0111028,
\newblock 

\bibitem{El-Khadra:1997mp}
A.~X. El-Khadra, A.~S. Kronfeld, and P.~B. Mackenzie,
\newblock Phys. Rev. {\bf D55}, 3933 (1997), hep-lat/9604004,
\newblock 

\bibitem{Kronfeld:2000ck}
A.~S. Kronfeld,
\newblock Phys. Rev. {\bf D62}, 014505 (2000), hep-lat/0002008,
\newblock 

\bibitem{Klassen:1998fh}
T.~R. Klassen,
\newblock Nucl. Phys. Proc. Suppl. {\bf 73}, 918 (1999), hep-lat/9809174,
\newblock 

\bibitem{Chen:2000ej}
P.~Chen,
\newblock Phys. Rev. {\bf D64}, 034509 (2001), hep-lat/0006019,
\newblock 

\bibitem{Okamoto:2001jb}
CP-PACS, M.~Okamoto {\em et~al.},
\newblock Phys. Rev. {\bf D65}, 094508 (2002), hep-lat/0112020,
\newblock 

\bibitem{Alford:2000an}
M.~G. Alford, I.~T. Drummond, R.~R. Horgan, H.~Shanahan, and M.~J. Peardon,
\newblock Phys. Rev. {\bf D63}, 074501 (2001), hep-lat/0003019,
\newblock 

\bibitem{Collins:2001pe}
UKQCD, S.~Collins {\em et~al.},
\newblock Phys. Rev. {\bf D64}, 055002 (2001), hep-lat/0101019,
\newblock 

\bibitem{Hein:2000qu}
J.~Hein {\em et~al.},
\newblock Phys. Rev. {\bf D62}, 074503 (2000), hep-ph/0003130,
\newblock 

\bibitem{Allton:1992zy}
UKQCD, C.~R. Allton {\em et~al.},
\newblock Phys. Lett. {\bf B292}, 408 (1992), hep-lat/9208018,
\newblock 

\bibitem{El-Khadra:1993ir}
A.~X. El-Khadra,
\newblock Nucl. Phys. Proc. Suppl. {\bf 30}, 449 (1993), hep-lat/9211046,
\newblock 

\bibitem{Choe:2001yg}
QCD-TARO, S.~Choe {\em et~al.},
\newblock Nucl. Phys. Proc. Suppl. {\bf 106}, 361 (2002), hep-lat/0110104,
\newblock 

\bibitem{Davies:1995db}
C.~T.~H. Davies {\em et~al.},
\newblock Phys. Rev. {\bf D52}, 6519 (1995), hep-lat/9506026,
\newblock 

\bibitem{Trottier:1997ce}
H.~D. Trottier,
\newblock Phys. Rev. {\bf D55}, 6844 (1997), hep-lat/9611026,
\newblock 

\bibitem{Shakespeare:1998dt}
N.~H. Shakespeare and H.~D. Trottier,
\newblock Phys. Rev. {\bf D58}, 034502 (1998), hep-lat/9802038,
\newblock 

\bibitem{Mathur:2002ce}
N.~Mathur, R.~Lewis, and R.~M. Woloshyn,
\newblock Phys. Rev. {\bf D66}, 014502 (2002), hep-ph/0203253,
\newblock 

\bibitem{Manke:1998qc}
CP-PACS, T.~Manke {\em et~al.},
\newblock Phys. Rev. Lett. {\bf 82}, 4396 (1999), hep-lat/9812017,
\newblock 

\bibitem{Orginos:1998fh}
K.~Orginos, W.~Bietenholz, R.~Brower, S.~Chandrasekharan, and U.~J. Wiese,
\newblock Nucl. Phys. Proc. Suppl. {\bf 63}, 904 (1998), hep-lat/9709100,
\newblock 

\bibitem{Hasenfratz:1998ft}
P.~Hasenfratz,
\newblock Nucl. Phys. Proc. Suppl. {\bf 63}, 53 (1998), hep-lat/9709110,
\newblock 

\bibitem{Richardson:1979bt}
J.~L. Richardson,
\newblock Phys. Lett. {\bf B82}, 272 (1979),
\newblock 

\bibitem{Allton:1998gi}
UKQCD, C.~R. Allton {\em et~al.},
\newblock Phys. Rev. {\bf D60}, 034507 (1999), hep-lat/9808016,
\newblock 

\bibitem{Bernard:2000gd}
C.~W. Bernard {\em et~al.},
\newblock Phys. Rev. {\bf D62}, 034503 (2000), hep-lat/0002028,
\newblock 

\bibitem{El-Khadra:2000zs}
A.~X. El-Khadra, S.~Gottlieb, A.~S. Kronfeld, P.~B. Mackenzie, and J.~N.
  Simone,
\newblock Nucl. Phys. Proc. Suppl. {\bf 83}, 283 (2000),
\newblock 

\bibitem{Stewart:2000ev}
C.~Stewart and R.~Koniuk,
\newblock Phys. Rev. {\bf D63}, 054503 (2001), hep-lat/0005024,
\newblock 

\bibitem{Grinstein:1996gm}
B.~Grinstein and I.~Z. Rothstein,
\newblock Phys. Lett. {\bf B385}, 265 (1996), hep-ph/9605260,
\newblock 

\bibitem{Gray:2002vk}
HPQCD, A.~Gray {\em et~al.},
\newblock (2002), hep-lat/0209022,
\newblock 

\bibitem{McNeile:2001cr}
UKQCD, C.~McNeile, C.~Michael, and K.~J. Sharkey,
\newblock Phys. Rev. {\bf D65}, 014508 (2002), hep-lat/0107003,
\newblock 

\bibitem{Morningstar:1999rf}
C.~J. Morningstar and M.~J. Peardon,
\newblock Phys. Rev. {\bf D60}, 034509 (1999), hep-lat/9901004,
\newblock 

\bibitem{Boyle:1997aq}
UKQCD, P.~Boyle,
\newblock Nucl. Phys. Proc. Suppl. {\bf 53}, 398 (1997),
\newblock 

\bibitem{Boyle:1998rk}
UKQCD, P.~Boyle,
\newblock Nucl. Phys. Proc. Suppl. {\bf 63}, 314 (1998), hep-lat/9710036,
\newblock 

\bibitem{Groom:2000in}
Particle Data Group, D.~E. Groom {\em et~al.},
\newblock Eur. Phys. J. {\bf C15}, 1 (2000),
\newblock 

\bibitem{Leinweber:1999ig}
D.~B. Leinweber, A.~W. Thomas, K.~Tsushima, and S.~V. Wright,
\newblock Phys. Rev. {\bf D61}, 074502 (2000), hep-lat/9906027,
\newblock 

\bibitem{Leinweber:2001ac}
D.~B. Leinweber, A.~W. Thomas, K.~Tsushima, and S.~V. Wright,
\newblock Phys. Rev. {\bf D64}, 094502 (2001), hep-lat/0104013,
\newblock 

\bibitem{Guo:2001ph}
X.~H. Guo and A.~W. Thomas,
\newblock Phys. Rev. {\bf D65}, 074019 (2002), hep-ph/0112040,
\newblock 

\bibitem{Becirevic:2001yh}
D.~Becirevic, V.~Lubicz, and G.~Martinelli,
\newblock Phys. Lett. {\bf B524}, 115 (2002), hep-ph/0107124,
\newblock 

\bibitem{Rolf:2002gu}
ALPHA, J.~Rolf and S.~Sint,
\newblock (2002), hep-ph/0209255,
\newblock 

\bibitem{Kronfeld:1998zc}
A.~S. Kronfeld,
\newblock Nucl. Phys. Proc. Suppl. {\bf 63}, 311 (1998), hep-lat/9710007,
\newblock 

\bibitem{Juge:2001dj}
K.~J. Juge,
\newblock Nucl. Phys. Proc. Suppl. {\bf 106}, 847 (2002), hep-lat/0110131,
\newblock 

\bibitem{El-Khadra:2002wp}
A.~X. El-Khadra and M.~Luke,
\newblock (2002), hep-ph/0208114,
\newblock 

\bibitem{Fritzsch:2001nv}
H.~Fritzsch and Z.-z. Xing,
\newblock Phys. Lett. {\bf B506}, 109 (2001), hep-ph/0102295,
\newblock 

\bibitem{Sint:1997jx}
S.~Sint and P.~Weisz,
\newblock Nucl. Phys. {\bf B502}, 251 (1997), hep-lat/9704001,
\newblock 

\bibitem{Gupta:1997sa}
R.~Gupta and T.~Bhattacharya,
\newblock Phys. Rev. {\bf D55}, 7203 (1997), hep-lat/9605039,
\newblock 

\bibitem{Kronfeld:2000gk}
A.~S. Kronfeld and J.~N. Simone,
\newblock Phys. Lett. {\bf B490}, 228 (2000), hep-ph/0006345,
\newblock 

\bibitem{Gockeler:1998fn}
M.~Gockeler {\em et~al.},
\newblock Phys. Rev. {\bf D57}, 5562 (1998), hep-lat/9707021,
\newblock 

\bibitem{Drummond:2002yg}
I.~T. Drummond, A.~Hart, R.~R. Horgan, and L.~C. Storoni,
\newblock (2002), hep-lat/0208010,
\newblock 

\bibitem{Sint:2000vc}
S.~Sint,
\newblock Nucl. Phys. Proc. Suppl. {\bf 94}, 79 (2001), hep-lat/0011081,
\newblock 

\bibitem{DiRenzo:2000nd}
F.~Di~Renzo and L.~Scorzato,
\newblock JHEP {\bf 02}, 020 (2001), hep-lat/0012011,
\newblock 

\bibitem{Martinelli:1995ty}
G.~Martinelli, C.~Pittori, C.~T. Sachrajda, M.~Testa, and A.~Vladikas,
\newblock Nucl. Phys. {\bf B445}, 81 (1995), hep-lat/9411010,
\newblock 

\bibitem{Chetyrkin:2000yt}
K.~G. Chetyrkin, J.~H. Kuhn, and M.~Steinhauser,
\newblock Comput. Phys. Commun. {\bf 133}, 43 (2000), hep-ph/0004189,
\newblock 

\bibitem{Capitani:1998mq}
ALPHA, S.~Capitani, M.~Luscher, R.~Sommer, and H.~Wittig,
\newblock Nucl. Phys. {\bf B544}, 669 (1999), hep-lat/9810063,
\newblock 

\bibitem{Anikeev:2001rk}
K.~Anikeev {\em et~al.},
\newblock (2001), hep-ph/0201071,
\newblock 

\bibitem{Shipsey:2002ye}
I.~Shipsey,
\newblock (2002), hep-ex/0207091,
\newblock 

\bibitem{Yamada:2001xp}
JLQCD, N.~Yamada {\em et~al.},
\newblock Nucl. Phys. Proc. Suppl. {\bf 106}, 397 (2002), hep-lat/0110087,
\newblock 

\bibitem{Kronfeld:2002ab}
A.~S. Kronfeld and S.~M. Ryan,
\newblock Phys. Lett. {\bf B543}, 59 (2002), hep-ph/0206058,
\newblock 

\bibitem{deDivitiis:1998kj}
UKQCD, G.~M. de~Divitiis {\em et~al.},
\newblock JHEP {\bf 10}, 010 (1998), hep-lat/9807032,
\newblock 

\bibitem{Abada:2002xe}
A.~Abada {\em et~al.},
\newblock (2002), hep-ph/0206237,
\newblock 

\bibitem{Anastassov:2001cw}
CLEO, A.~Anastassov {\em et~al.},
\newblock Phys. Rev. {\bf D65}, 032003 (2002), hep-ex/0108043,
\newblock 

\bibitem{Artuso:2002zh}
M.~Artuso and E.~Barberio,
\newblock (2002), hep-ph/0205163,
\newblock 

\bibitem{Luke:1990eg}
M.~E. Luke,
\newblock Phys. Lett. {\bf B252}, 447 (1990),
\newblock 

\bibitem{Bernard:1993ey}
C.~W. Bernard, Y.~Shen, and A.~Soni,
\newblock Phys. Lett. {\bf B317}, 164 (1993), hep-lat/9307005,
\newblock 

\bibitem{Booth:1994zb}
UKQCD, S.~P. Booth {\em et~al.},
\newblock Phys. Rev. Lett. {\bf 72}, 462 (1994), hep-lat/9308019,
\newblock 

\bibitem{Bowler:1995bp}
UKQCD, K.~C. Bowler {\em et~al.},
\newblock Phys. Rev. {\bf D52}, 5067 (1995), hep-ph/9504231,
\newblock 

\bibitem{Bowler:2002zh}
UKQCD, K.~C. Bowler, G.~Douglas, R.~D. Kenway, G.~N. Lacagnina, and C.~M.
  Maynard,
\newblock Nucl. Phys. {\bf B637}, 293 (2002), hep-lat/0202029,
\newblock 

\bibitem{Briere:2002ew}
CLEO, R.~A. Briere {\em et~al.},
\newblock Phys. Rev. Lett. {\bf 89}, 081803 (2002), hep-ex/0203032,
\newblock 

\bibitem{Caprini:1998mu}
I.~Caprini, L.~Lellouch, and M.~Neubert,
\newblock Nucl. Phys. {\bf B530}, 153 (1998), hep-ph/9712417,
\newblock 

\bibitem{Boyd:1997kz}
C.~G. Boyd, B.~Grinstein, and R.~F. Lebed,
\newblock Phys. Rev. {\bf D56}, 6895 (1997), hep-ph/9705252,
\newblock 

\bibitem{Hashimoto:2001nb}
S.~Hashimoto, A.~S. Kronfeld, P.~B. Mackenzie, S.~M. Ryan, and J.~N. Simone,
\newblock Phys. Rev. {\bf D66}, 014503 (2002), hep-ph/0110253,
\newblock 

\bibitem{Bowler:1998ej}
UKQCD, K.~C. Bowler {\em et~al.},
\newblock Phys. Rev. {\bf D57}, 6948 (1998), hep-lat/9709028,
\newblock 

\bibitem{Jones:1998ubw}
B.~D. Jones and R.~M. Woloshyn,
\newblock Phys. Rev. {\bf D60}, 014502 (1999), hep-lat/9812008,
\newblock 

\bibitem{Shanahan:1999mv}
UKQCD, H.~P. Shanahan, P.~Boyle, C.~T.~H. Davies, and H.~Newton,
\newblock Phys. Lett. {\bf B453}, 289 (1999), hep-lat/9902025,
\newblock 

\bibitem{Stone:2001jh}
S.~Stone,
\newblock (2001), hep-ph/0112008,
\newblock 

\bibitem{Wise:2001ii}
M.~B. Wise,
\newblock (2001), hep-ph/0111167,
\newblock 

\bibitem{Ishizuka:2002nm}
N.~Ishizuka,
\newblock (2002), hep-lat/0209108,
\newblock 

\bibitem{Bernard:1990ii}
C.~W. Bernard, J.~Simone, and A.~Soni,
\newblock Nucl. Phys. Proc. Suppl. {\bf 17}, 504 (1990),
\newblock 

\bibitem{Abada:1990ns}
European Lattice, A.~Abada {\em et~al.},
\newblock Nucl. Phys. Proc. Suppl. {\bf 17}, 518 (1990),
\newblock 

\bibitem{Bernard:1991dp}
C.~W. Bernard, J.~N. Simone, and A.~Soni,
\newblock Nucl. Phys. Proc. Suppl. {\bf 20}, 434 (1991),
\newblock 

\bibitem{Michael:1989mf}
C.~Michael,
\newblock Nucl. Phys. {\bf B327}, 515 (1989),
\newblock 

\bibitem{Maiani:1990ca}
L.~Maiani and M.~Testa,
\newblock Phys. Lett. {\bf B245}, 585 (1990),
\newblock 

\bibitem{Lellouch:2000pv}
L.~Lellouch and M.~Luscher,
\newblock Commun. Math. Phys. {\bf 219}, 31 (2001), hep-lat/0003023,
\newblock 

\bibitem{Lin:2001ek}
C.~J.~D. Lin, G.~Martinelli, C.~T. Sachrajda, and M.~Testa,
\newblock Nucl. Phys. {\bf B619}, 467 (2001), hep-lat/0104006,
\newblock 

\bibitem{Ciuchini:1996mq}
M.~Ciuchini, E.~Franco, G.~Martinelli, and L.~Silvestrini,
\newblock Phys. Lett. {\bf B380}, 353 (1996), hep-ph/9604240,
\newblock 

\end{thebibliography}

%

\end{document}